\documentclass[cmp]{svjour}
\usepackage{latexsym}
\usepackage{graphics}
\usepackage{amsmath,amssymb}

\def\bbbz{\Bbb{Z}}
\def\bbbc{\Bbb{C}}
\def\bbbr{\Bbb{R}}
\def\bbbt{\Bbb{T}}
\def\bbbs{\Bbb{S}}

\def\ad{\mbox{ad\,}}
\def\tr{\mbox{tr\,}}
\def\im{\mbox{Im\,}}
\def\diag{\mbox{diag\,}}

\def\openone{\leavevmode\hbox{\small1\kern-3.3pt\normalsize1}}

\def\f{\boldsymbol f}
\def\bchi{\boldsymbol \chi }
\allowdisplaybreaks

\begin{document}

\title{How many types of soliton solutions do we know?}
\author{V. S. Gerdjikov\inst{1}\and D. J. Kaup\inst{2}
\and N. A. Kostov\inst{1}\and T. I. Valchev\inst{1}}

\institute{Institute for Nuclear Research and Nuclear Energy,\\
Bulgarian Academy of Sciences, \\ 72 Tsarigradsko chaussee,
1784 Sofia, BULGARIA \and Department of Mathematics,
University of Central Florida\\ Orlando, Florida 32816-1364, USA}

\date{Received: date / Accepted: date}
% Add name of the expert who has communicated your paper
\communicated{name}
\maketitle
\begin{abstract}
We consider several ways of how one could classify the various types of
soliton solutions related to nonlinear evolution equations which are
solvable by the inverse scattering method. In doing so we make use
of the fundamental analytic solutions, the dressing procedure,
the reduction technique and other tools characteristic for that method.
\end{abstract}

\section{Introduction}
\label{intro}

It is our impression that the question in the title has not been answered
satisfactorily even for some of the best known type of soliton equations
such as the $N $-wave equations, the multicomponent  NLS equations and
others.

We use the term `soliton solution' as a special solution to a given
nonlinear evolution equation (NLEE) which is solvable by the inverse
scattering method \cite{ZMNP,FaTa}. That means that the NLEE allows
Lax representation:
\[[L(\lambda ),M(\lambda )]=0,\]
where $L(\lambda ) $ and $M(\lambda ) $ are two linear differential
operators. In what follows we take them to be first order matrix
differential operators
\begin{eqnarray*}
L\psi (x,t,\lambda ) &\equiv & i \partial_x\psi  +U(x,t,\lambda )
\psi (x,t,\lambda ), \\
M\psi (x,t,\lambda ) &\equiv & i \partial_t\psi +V(x,t,\lambda ) \psi
(x,t,\lambda ).
\end{eqnarray*}

The one-soliton solutions are related to one or a set of several
discrete eigenvalues of the Lax operator $L $. Therefore one  first has to
study the different  configurations of discrete eigenvalues of $L $, see
\cite{Harnad1}. The next step in classifying the types of one-soliton
solutions is related to the study of their internal degrees of freedom.

In order to make the problem not too difficult we will specify $L
$ to be the generalized Zakharov-Shabat system:
\[L(\lambda )\psi (x,\lambda ) \equiv i\partial_x\psi + (q(x) -
\lambda J) \psi (x,\lambda ) =0.\]
where we take the potential $q(x,t) $ to be $n\times n $
matrix-valued smooth function of $x $ tending to zero
sufficiently rapid as $x\to\pm\infty $. We also restrict $J $ to
be a real constant diagonal matrix with different eigenvalues.

We will try to answer the question in the title first for the simplest
class of Lax operators of Zakharov-Shabat type with real-valued $J$.
In doing this we will be using the dressing method, one of the best known
methods for constructing reflectionless potentials and soliton solutions.
This paper is intended as a natural continuation of the work \cite{Kaup-Ger}
published several years ago by two of the authors.

In Section \ref{sec:2} below we first outline the well known facts
about the soliton types of NLEE solvable by the $sl(2) $
Zakharov-Shabat system. In Section \ref{sec:3} we treat the
different one-soliton solutions for the $sl(n) $ Zakharov-Shabat
systems related to the subalgebras $sl(p) $. Most of our results
are illustrated for the $sl(5) $ system, but it is not difficult
to extend them to any $sl(n) $ system. The structure of the
eigenfunctions of $L(\lambda ) $ corresponding to the different
types of solitons is outlined in Section \ref{sec:4}. In the last
Section we discuss possible generalizations to other
Zakharov-Shabat systems having additional symmetry properties. The
presence of symmetries modifies the spectrum of the scattering
operator $L$. A typical example is the reduction of the type
$U^{\dag}(x,t,\lambda^{\ast})=U(x,t,\lambda)$. As a result the
eigenvalues are pairwise symmetrical with respect to the real axis
$\mathbb{R}$, that is $\lambda^{+}=(\lambda^{-})^{\ast}$ holds
true. The soliton solutions are connected with two eigenvalues
(doublet solitons). Another common situation is when we have a
$\mathbb{Z}_h$ type reduction. The continuous spectrum of the
Lax operator which is compatible with that reduction consists of
$2h$ rays \cite{cake}. Thus the complex $\lambda$-plane is split
into identical sectors which possess equal number of eigenvalues.
The soliton solution is associated with a multiplet of discrete
eigenvalues (multiplet solitons).

\section{Preliminaries}\label{sec:1}

In this section we shall outline some basic features
of the mathematical machinery we are about to use for
the classification of soliton solutions.

Integrability or more precisely S-integrability of a NLEE
means that the NLEE can be presented as a zero curvature
condition
\begin{equation}\label{eq:Lax}
[L(\lambda ),M(\lambda )]=0,
\end{equation}
of two first order linear  matrix differential operators
$L(\lambda ) $ and $M(\lambda )$ of the form
\begin{eqnarray}\label{eq:L-}
L\psi (x,t,\lambda ) &\equiv & i \partial_x\psi  +U(x,t,\lambda )
\psi (x,t,\lambda ) =0, \\
\label{eq:M-}
M\psi (x,t,\lambda ) &\equiv & i \partial_t\psi +V(x,t,\lambda ) \psi
(x,t,\lambda ) = \psi (x,t,\lambda )C(\lambda ).
\end{eqnarray}
The potentials $U(x,t,\lambda)$ and $V(x,t,\lambda)$ are typically
chosen as elements of some semismple Lie algebra $g$ (the
fundamental solutions $\psi(x,t,\lambda)$ belong to the
corresponding Lie group $G$). We shall mainly deal with the
algebra $sl(n)$.

\begin{remark}
The compatibility condition (\ref{eq:Lax}) means that the Lax
operators $L$ and $M$ possess the same eigenfunctions. The matrix
$C(\lambda)$ depends on the definition of Jost solutions.
\end{remark}

The compatibility condition (\ref{eq:Lax}) which must hold true
identically with respect to $\lambda $ takes the form
\begin{equation}\label{eq:C-}
i \partial_x V -i \partial_t U
+[U(x,t,\lambda ), V(x,t,\lambda )]=0
\end{equation}
and it is valid for any choice of $C(\lambda )$. For simplicity
we shall resrict our considerations on scattering operators
of Zakharov-Shabat type (GZS)
\begin{equation}\label{eq:ZS}
L(\lambda )\psi (x,t,\lambda ) \equiv i\partial_x\psi + (q(x,t) -
\lambda J) \psi (x,t,\lambda ) =0.
\end{equation}
The matrix $J$ is a real traceless diagonal matrix, i.e. a real
Cartan element of $sl(n)$, while $q(x,t)$ is a matrix with zero diagonal elements.
Since $J$ is a real matrix one can introduce an ordering of its elements
$J_1>J_2>\cdots >J_n$. By carrying out a gauge transformation which
commutes with $J $, we can always take $q(x,t) $ to be of the form
$q(x,t)=[J,Q(x,t)] $, i.e. $q_{jj}\equiv 0 $. The linear subspace in $sl(n)$
of matrix-valued functions $q(x,t)=[J,Q(x,t)]$ are known in literature
to be the co-adjoint orbit of $\mathfrak{g}$ passing through $J $.
The co-adjoint orbits can be supplied in a natural way with a non-degenerate
symplectic structure which makes them natural choices for the phase spaces
$\mathcal{M}_J $ and Hamiltonian structures of the corresponding NLEE.

The class of NLEE related to $L(\lambda )$ are systems of
equations for the functions $Q_{jk}(x,t) $, which may be written
in the compact form \cite{ZaMa,DJK0,VGPK,LMP5}:
\begin{equation}\label{eq:NLEE}
i\partial_t Q + 2 \sum_{k=1}^{4} \Lambda^k[H_k,
Q(x,t)]=0,
\end{equation}
where $H_k$, $ \tr H_k=0 $ are constant diagonal matrices and $f(\lambda)
=\sum_{k=1}\lambda^k H_k$ is the dispersion law of the NLEE. Here and
below we define:
\begin{equation}\label{eq:adJ}
(\ad_J X)_{ks}\equiv ([J, X])_{ks} = (J_k-J_s)X_{ks}, \qquad
\left(\ad_J^{-1} X\right)_{ks} = \frac{X_{ks}}{ J_k-J_s } ,
\end{equation}
for all $X\in \mathcal{M}_J $, i.e. $X_{kk}=0 $.
The operator $\Lambda$ is either one of the recursion  operators
$\Lambda_\pm $, acting on the space $\mathcal{M}_J$ of $n\times n $
off-diagonal matrix-valued functions as follows
\begin{equation}\label{eq:Lam}
\Lambda _{\pm} X \equiv \ad_J^{-1}\left( i \partial_x X +
P_0[q(x),X(x)]+ i \sum_{k=1}^{5} [Q(x),E_{k,k}] \int_{\pm\infty }^{x}
\mathrm{d} y\; \tr\, \left( Q(y),  X(y)], E_{k,k}\right) \right).
\end{equation}
where $P_0\cdot $ is the projector $\ad_J^{-1}\ad_J\cdot $.
Choosing $H_1=I=\diag (I_1,\dots ,I_n)\neq\openone $, so that the dispersion law
$f(\lambda )=-\lambda I $ is a linear function of $\lambda $  we get a
system, generalizing the well known $N $-wave equation:
\begin{equation}\label{eq:N-w0}
i[J,Q_t] -i[I,Q_x] - \left[ [J,Q],[I,Q]\right]=0,
\end{equation}
which contains  $N=n(n-1)$ complex-valued functions $Q_{ij}(x,t) $.

In order to describe the soliton solutions we shall use the so-called dressing
procedure \cite{ZS-dress}. For that purpose we need some basic facts on the direct
scattering problem of $L$ operator.

Let $\psi_{\pm}(x,t,\lambda)$ are two fundamental solutions of the GZS system
(\ref{eq:ZS}). If they satisfy the requirement
\begin{equation}\label{jost}
\lim_{x\to\pm\infty}\psi_{\pm}(x,t,\lambda)e^{i\lambda Jx}=\openone
\end{equation}
they shall be called Jost solutions. The Jost solutions are
interrelated via
\begin{equation}\label{T-matrix}
\psi_-(x,t,\lambda)=\psi_+(x,t,\lambda)T(t,\lambda),
\end{equation}
where $T(t,\lambda)$ is called a scattering matrix. The scattering
matrix is $x $-indepedent  and its time evolution is driven by the
linear equation
\begin{equation}\label{eq:T_t}
i\partial_t T+[f(\lambda),T]=0.
\end{equation}
For the case of the $N$-wave equation we have
\begin{equation}\label{eq:T_t_waves}
i \partial_t T  - \lambda \left[ I, T(\lambda ,t)\right]=0.
\end{equation}
Thus, if $Q(x,t) $ satisfies the $N $-wave system (\ref{eq:N-w0}) we get:
\begin{equation}\label{eq:dT/dt}
\partial_t T_{kk} (\lambda )=0, \qquad i\partial_t T_{jk}
(\lambda ) - \lambda (I_j -I_k)T_{jk}(\lambda,t )
=0.
\end{equation}
The set of matrix elements of $T(\lambda ,t) $ must satisfy a number of
relations. Indeed, they are uniquely determined by $Q(x,t) $, i.e. by
$n(n-1) $ complex functions of $x $, so it seems natural that there
shouldn't be more that $n(n-1) $ independent functions among
$T_{jk}(\lambda ) $ for $\lambda $ on the real axis. Of course $T(\lambda
,t) $ must  satisfy the `unitarity' condition $\det T(\lambda,t )=1 $. The
rest of these relations follow from the analyticity properties of certain
combinations of matrix elements of $T(\lambda ,t) $. These analyticity
properties must follow naturally from the corresponding fundamental analytic
solutions (FAS) $\chi^\pm(x,t,\lambda )$.

The Jost solutions are well defined only for $\lambda\in\bbbr$, i.e. they do not
have necessarily analytic properties beyond the real axis. This can be seen easily
if one reformulates the problem (\ref{eq:ZS}) in terms of a Volterra type integral
equation
\begin{equation}
\xi_{\pm}(x,t,\lambda)=\openone+i\int^x_{\pm\infty}\mathrm{d}y
e^{i\lambda J(y-x)}q(y,t)\xi_{\pm}(y,t,\lambda)e^{i\lambda J(x-y)},
\end{equation}
where $\xi_{\pm}(x,t,\lambda)=\psi_{\pm}(x,t,\lambda)e^{i\lambda Jx}$ represents
another set of fundamental solutions but this time to the linear problem
\[i\partial_x\xi(x,t,\lambda) + q(x,t)\xi(x,t,\lambda) -
\lambda [J, \xi (x,t,\lambda )] =0.\]
It is easy to see that only the first and the last columns of $\psi_{+}(x,t,\lambda )$
and $\psi_{-} (x,t,\lambda )$ allow analytic extensions in $\lambda$ off the real
axis; generally the other columns {\em do not} have analyticity properties. Nevertheless it is
again possible to introduce FAS \cite{Sh,ZMNP}. Taking into account the ordering
introduced above one is able to construct new fundamental solutions
\begin{equation}
\xi^{\pm}_{kl}(x,t,\lambda)=\left\{\begin{array}{cc}\delta_{kl}
+i\int^x_{\pm\infty}\mathrm{d}y e^{i\lambda(J_k-J_l)(y-x)}
(q\xi^{\pm})_{kl}(y,t,\lambda),& k\leq l,\\
i\int^x_{\mp\infty}\mathrm{d}y e^{i\lambda(J_k-J_l)(y-x)}
(q\xi^{\pm})_{kl}(y,t,\lambda),& k>l.\end{array}\right.
\end{equation}
to possess analytic properties in the half planes $\bbbc_{\pm}$
of the spectral parameter. This definition can be rewritten
using Gauss factors of the scattering matrix $T$
\begin{equation}\label{fas_jost}
\chi^{\pm}(x,t,\lambda)=\psi_{-}(x,t,\lambda)\mathbb{S}^{\pm}(t,\lambda)
=\psi_{+}(x,t,\lambda)\mathbb{T}^{\mp}(t,\lambda),
\end{equation}
where $T(t,\lambda)=\mathbb{T}^{\mp}(t,\lambda)(\mathbb{S}^{\pm}(t,\lambda))^{-1}$
and $\chi^{\pm}(x,t,\lambda)=\xi^{\pm}(x,t,\lambda)e^{-i\lambda Jx}$.
The matrix elements of $\bbbt^\pm(\lambda )$ and $\bbbs^\pm(\lambda )  $
can be expressed in terms of the minors of $T(\lambda ) $. Here we note
that their diagonal elements can be given by:
\begin{eqnarray}\label{eq:ST-dia}
\bbbs_{jj}^{+}(\lambda ) = m_{j-1}^{+}(\lambda ), \qquad
\bbbt_{jj}^{-}(\lambda ) = m_{j}^{+}(\lambda ), \\
\bbbt_{jj}^{+}(\lambda ) = m_{n-j}^{-}(\lambda ), \qquad
\bbbs_{jj}^{-}(\lambda ) = m_{n+1-j}^{-}(\lambda ),
\end{eqnarray}
where $m_0^\pm =m_n^\pm =1 $ and by $m_k^+(\lambda ) $ (resp.
$m_k^-(\lambda ) $) we have denoted the upper (resp. lower) principal
minors of $ T(\lambda ) $ of order $k $, e.g.:
\begin{eqnarray}\label{eq:mk}
m_k^+(\lambda ) &=& \left\{ \begin{array}{cccc} 1 & 2 & \hdots &
k\\ 1 & 2 &\hdots & k\end{array}\right\}, \qquad k=1,\hdots,5\\
m_k^-(\lambda ) &=& \left\{ \begin{array}{cccc} 5-k+1 & 5-k+2 & \hdots & 5\\
5-k+1 & 5-k+2 & \hdots & 5 \end{array}
\right\}_{T(\lambda )},\\
\left\{\begin{array}{cccc}
i_1 & i_2 & \hdots & i_k\\
j_1 & j_2 & \hdots & j_k
\end{array}\right\}_{T(\lambda)}& \equiv &
\det\left(\begin{array}{cccc}
T_{i_1j_1} & T_{i_1j_2} & \hdots & T_{i_1j_k}\\
T_{i_2j_1} & T_{i_2j_2} & \hdots & T_{i_2j_k}\\
\vdots     & \vdots     & \ddots & \vdots    \\
T_{i_kj_1} & T_{i_kj_2} & \hdots & T_{i_kj_k}
\end{array}\right).
\end{eqnarray}

As a consequence of the analyticity of the FAS, it follows that
the minors $m_k^+(\lambda ) $  (resp. $m_k^-(\lambda ) $)  are analytic
functions for $\lambda \in \bbbc_+$ (resp. for  $\lambda \in \bbbc_-$).

One can construct the kernel of the resolvent of $L(\lambda ) $ in terms
of the FAS \cite{VGPK,LMP5} from which it follows that the resolvent
has poles for all $\lambda _k^\pm $ which happen to be zeroes of any of
the minors $m_k^\pm(\lambda ) $. Therefore what we have now is that
each of the minors $m_k^\pm(\lambda ) $ may be considered to be an analog
of the Evans function, and thus now, there is more than one Evans
function.

There exist different methods to solve a NLEE possessing a Lax representation
: Gel'fand-Levitan-Marchenko integral equation, Hirota method, dressing method etc.
We shall use the dressing Zahkarov-Shabat method \cite{ZS-dress}. Let $\psi_0(x,t,\lambda)$
be a fundamental solution of Zakharov-Shabat's system with a known potential
$U_0(x,t,\lambda)=q_0(x,t)-\lambda J$. Consider a new function
$\psi(x,t,\lambda)=u(x,t,\lambda)\psi_0(x,t,\lambda)$ which is a solution to a
Zakharov-Shabat's problem with some potential $q(x,t)-\lambda J$
to be found. This requires that $u(x,t,\lambda)$ satisfies
\begin{equation}\label{dress_pde}
i\partial_x u+qu-uq_0-\lambda [J,u]=0.
\end{equation}
The dressing procedure transforms the Jost solutions $\psi_{\pm,0}(x,t\lambda)$,
the scattering matrix $T_0(t,\lambda)$ and the fundamental solution
$\chi^{\pm}_{0}(x,t,\lambda)$ of the generalized Zakharov-Shabat system with
a potential $U_0(x,t,\lambda)$ in the following fashion
\begin{eqnarray}\label{jost_dress}
\psi_{\pm}(x,t,\lambda)& = & u(x,t,\lambda)\psi_{\pm,0}(x,t,\lambda)u^{-1}_{\pm}(\lambda),\\
T(t,\lambda)& = & u_{+}(\lambda)T_0(t,\lambda)u^{-1}_{-}(\lambda),\label{T_dress}\\
\chi^{\pm}(x,t,\lambda)&=&u(x,t,\lambda)\chi^{\pm}_0(x,t\lambda)u^{-1}_{-}(\lambda)\label{FAS_dress}.
\end{eqnarray}
The normalizing factors $u_{\pm}(\lambda)=\lim_{x\to\pm\infty}u(x,t,\lambda)$ ensures
the proper assymptotics of the dressed solutions $\psi_{\pm}(x,t,\lambda)$.

Zakharov and Shabat \cite{ZMNP} proposed the following ansatz for
the dressing factor $u(x,t,\lambda)$
\begin{equation}\label{dresfac}
u(x,t,\lambda)=\openone+(c(\lambda)-1)P(x,t), \qquad
c(\lambda)=\frac{\lambda-\lambda^{+}}{\lambda-\lambda^{-}},
\end{equation}
where $P$ is a projector ($P^2=P$) which can be expressed via the fundamental
analytic solutions (FAS) and $\lambda^{+}$ (resp. $\lambda^-$) is an arbitrary
complex number in the upper (resp. lower) half plane $\bbbc_{+}$ (resp.
$\bbbc_{-}$). In the simplest case when $\mathrm{rank}\,P=1$ it reads
\begin{equation}\label{dresfac_1}
P(x,t)=\frac{|n(x,t)\rangle\langle m(x,t)|}{\langle m(x,t)|n(x,t)\rangle},
\end{equation}
where
\begin{equation} |n(x,t)\rangle=\chi^+_0(x,t,\lambda^+)|n_0\rangle,
\qquad\langle m(x,t)|=\langle m_0|(\chi_0^-(x,t,\lambda^-))^{-1}.
\label{vec2}\end{equation}
By taking the limit $\lambda\to\infty$ in equation (\ref{dress_pde}) we
obtain an interrelation between the seed solution $q_0$ and the new one
\begin{equation}\label{q0_to_q}
q=q_0+(\lambda^{-}-\lambda^{+})[J,P].
\end{equation}
Thus starting from a known solution of the NLEE we can find another
solution by simply dressing it with some factor $u(x,t,\lambda)$. An
important particular case is when $q_0=0$. The dressed solution is called
a 1-soliton solution. The fundamental analytic solution in the soliton
case is given by a plane wave $\chi^{\pm}_0(x,t)=e^{-i\lambda^+(Jx+It)}$.
Repeating the same procedure one derives step by step the multisoliton
solution of the corresponding NLEE, i.e.
\begin{equation}
0\to q^{1s}\to q^{2s}\to\hdots\to q^{ns}.
\end{equation}

Many integrable equations correspond to Lax operators that obey
some additional symmetry conditions of algebraic nature. That is
why it is worthwhile to outline some aspects of the theory of such
Lax operators.

Let an action of a discrete group $G_R$ to be referred to as a
reduction group be given on the set of fundamental solutions to
the generalized Zakharov-Shabat system (\ref{eq:ZS}) as follows
\begin{equation}
\tilde{\psi}(x,\lambda)=K\psi(x,k(\lambda))K^{-1},
\end{equation}
where $k:\bbbc\to\bbbc$ is a conformal map. This action yields
another action on the potential in the scattering operator $L$
\begin{equation}
KU(x,k(\lambda))K^{-1}=U(x,\lambda)
\end{equation}
A common case is when $G_R=\bbbz_2$. Then the action of $\bbbz_2$ might
involve external automorphisms of $SL(n)$ as well
\begin{eqnarray}
\tilde{\psi}(x,\lambda)=K\left(\psi^T(x,k_1(\lambda))\right)^{-1}K^{-1},& \Rightarrow&
KU^{T}(x,k_1(\lambda))K^{-1}=-U(x,\lambda),\label{sym1}\\
\tilde{\psi}(x,\lambda)=K\psi^{\ast}(x,k_2(\lambda))K^{-1},& \Rightarrow&
KU^{\ast}(x,k_2(\lambda))K^{-1}=-U(x,\lambda).\label{sym2}
\end{eqnarray}
In particular, if $\bbbz_2$ acts trivially on the complex plane of the spectral parameter
, i.e. $k=id$, then the symmetry condition (\ref{sym1}) resricts the potential $U(x,\lambda)$
to a certain subalgebra of $sl(n)$. For example, suppose $K^T=K$ then $U(x,\lambda)$
belongs to the orthogonal algebra $so(n)$. The existence of a $\bbbz_2$ reduction requires a
modification of the dressing factor $u(x,t,\lambda)$ as follows
\begin{equation}\label{u_orthogon}
u(x,t,\lambda)=\openone+\left(\frac{1}{c(\lambda)}-1\right)P(x,t)+
\left(c(\lambda)-1\right)\overline{P}(x,t),
\end{equation}
where $P(x,t)$ is a projector of rank 1 and
\begin{equation}
\overline{P}(x,t)=KP^T(x,t)K^{-1}.
\end{equation}
The projector itself can be expressed through the FAS $\chi^{\pm}(x,t,\lambda)$
\begin{equation}
P(x,t)=\frac{|n(x,t)\rangle \langle m(x,t)|}{\langle m(x,t)| n(x,t)\rangle},
\quad |n(x,t)\rangle =\chi(x,t,\lambda^-)|n_0\rangle,
\quad \langle m(x,t)| =\langle m_0| (\chi(x,t,\lambda^+))^{-1}.
\end{equation}

\section{Zakharov-Shabat system and $sl(2) $ solitons}\label{sec:2}

The best known examples of NLEE are related to the Zakharov-Shabat system
which is associated with the $sl(2)$ algebra as follows
\begin{equation}\label{eq:U_0}
L\psi(x,t,\lambda)\equiv (i\partial_x+q(x,t) -\lambda \sigma _3)\psi(x,t,\lambda),
\qquad q(x,t)= q^{+}\sigma_{+}+q^{-}\sigma_{-}=
\left(\begin{array}{cc} 0 & q^+ \\
q^- & 0 \end{array}\right),
\end{equation}
where $\sigma_{\pm}=(\sigma_1\pm i\sigma_2)/2$ and $\sigma_1$, $\sigma_2$
and $\sigma_3$ are the Pauli matrices.

The class of NLEE for the functions $q^\pm(x,t)$ related to (\ref{eq:U_0})
can be written in the compact form
\cite{AKNS,KN78,GeHr}:
\begin{equation}\label{eq:NLEE0}
i \sigma _3\partial_t q  + 2
f(\Lambda)q(x,t)=0,
\end{equation}
where $f(\lambda )$ is the dispersion law of the NLEE and
$\Lambda$ is one of the recursion operators, acting on the space
$\mathcal{M}_0$ of off-diagonal matrix-valued functions as
follows:
\begin{eqnarray}\label{eq:Lam_0}
\Lambda _{\pm} X\equiv \frac{i}{ 4 } \left[ \sigma _3 ,
\partial_x X\right]
 +  \frac{i}{2} q(x) \int_{\pm\infty }^{x}
\mathrm{d} y \, \tr\, \left( q(y), [\sigma _3, X(y)]\right).
\end{eqnarray}
The simplest nontrivial example of NLEE is related to a dispersion law
of the type $f(\lambda ) = -2\lambda ^2$. This is the nonlinear Schr\"odinger
equation
\begin{eqnarray}\label{eq:1.35}
&& i q_{t}^+ + q_{xx}^+ + 2(q^{+}(x,t))^2q^{-}(x,t) = 0
,\nonumber
\\ && i q_{t}^- - q_{xx}^- - 2(q^-(x,t))^2q^+(x,t) = 0.
\end{eqnarray}
Another well known example is provided by a cubic dispersion law
$f(\lambda ) = 4\lambda ^3$, one gets the system
\begin{eqnarray}\label{eq:1.37}
&& q_{t}^+ + q_{xxx}^+ + 6q^+(t)q^-(x,t)q_{x}^+ = 0
,\nonumber\\ && q_{t}^- + q_{xx}^- +
6q^-(x,t)q^+(x,t)q_{x}^- = 0 .
\end{eqnarray}
directly linking to the Korteweg de Vries equation.

As we discussed in the previous section the scattering theory is based
on introducing Jost solutions of $L(\lambda )$, scattering matrix, fundamental
solutions etc. In the $sl(2)$ case  the Jost solutions are $2\times
2 $ matrix-valued solutions defined by an analog of (\ref{jost})
where the matrix $J$ is simply substituted by $\sigma_3$.
Then one introduces the scattering matrix $T(\lambda ,t) $ by:
\begin{equation}\label{eq:T_0}
T(\lambda ,t) \equiv (\psi_{+}(x,t,\lambda ))^{-1}\psi_{-} (x,t,\lambda ) =
\left( \begin{array}{cc} a^+(\lambda ) & -b^-(\lambda,t ) \\
b^+(\lambda,t ) & a^-(\lambda ) \end{array}\right),
\end{equation}
which is $x $-indepedent. The $t $-dependence of the scattering matrix is
driven by
\begin{equation}\label{eq:T_t0}
i \partial_t T + \left[ f(\lambda )\sigma _3, T(\lambda
,t)\right]=0,
\end{equation}
Thus, if $q^\pm (x,t) $ satisfy the system of equations (\ref{eq:NLEE0})
we get
\begin{equation}\label{eq:dT_0dt}
\partial_t a^\pm (\lambda ) =0, \qquad i \partial_t b^\pm
(\lambda ) \mp 2f(\lambda )b^\pm(\lambda ) =0,
\end{equation}

The matrix elements of $T(\lambda ,t) $ are not independent. They
satisfy the `unitarity' condition $\det T(\lambda )\equiv a^+a^- +
b^+b^-=1 $. Besides the diagonal elements $a^+ $ and $a^- $ allow
analytic extension with respect to $\lambda  $ in the upper and lower
complex $\lambda  $-plane respectively. In fact the minimal set of
scattering data which uniquely determines both the scattering matrix and
the corresponding potential $q(x) $ consists of two types of variables:
i)~the reflection coefficients $\rho^\pm(\lambda )=b^\pm/a^\pm $
defined for real $\lambda \in \bbbr $ and ii)~a discrete set of scattering
data including the discrete eigenvalues $\lambda _k^\pm \in\bbbc_\pm $ and
the constants $C_k^\pm $ which determine the norm of the corresponding
Jost solutions \cite{DJK}.

A simple analysis shows that the first column of $\psi_{+}$ allows analytic
continuation in the lower half plane of the spectral parameter while the last
one --- in the upper half plane (for $\psi_{-}$ the opposite holds true)
\begin{equation}
\psi_{+} (x,t,\lambda ) = [\psi_{+} ^-,\psi_{+} ^+], \qquad
\psi_{-} (x,t,\lambda ) = [\psi_{-} ^+, \psi_{-}^-].
\end{equation}
The superscripts $\pm $ in the columns of the Jost solutions refer
to their analyticity properties while the subscripts $\pm$ refer
to different Jost solutions (with different limits of $x$).
The fundamental analytic solutions are constructed in the following manner
\begin{eqnarray}\label{eq:FAS-0}
\chi ^+(x,t,\lambda ) = [\psi_{-} ^+,\psi_{+} ^+], \qquad
\chi^-(x,t,\lambda ) = [\psi_{+} ^-,\psi_{-} ^-].
\end{eqnarray}

The functions $a^\pm(\lambda )=\det \chi ^\pm(x,\lambda ) $ are
known as the Evans functions \cite{ZaSh,Eva1} of the system
$L(\lambda ) $. Their importance comes from the fact that they
are $t $-independent (see eq. (\ref{eq:dT_0dt})), and therefore
they (or rather $\ln a^\pm $) can be viewed as generating
functionals of the (local) integrals of motion. In addition it is
known that their zeroes determine the discrete eigenvalues of
$L(\lambda ) $:
\begin{equation}\label{eq:dis-eig}
a^+(\lambda_k^+)=0, \qquad \lambda _{k}^+ \in \bbbc_+; \qquad
a^-(\lambda _{k}^-)=0, \qquad \lambda _{k}^- \in \bbbc_-.
\end{equation}

One can define the soliton solutions of the NLEE as the ones for which
$\rho^\pm(\lambda )=0 $ for all $\lambda\in\bbbr$. Thus the soliton
solutions of the NLEE (\ref{eq:NLEE0}) are parametrized by the discrete
eigenvalues and the constants $C_{k}^\pm $ whose $t $-dependence is
determined from
\begin{equation}\label{eq:dC/dt}
\frac{\mathrm{d}\lambda _{k}^\pm}{\mathrm{d}t} =0, \qquad i\frac{\mathrm{d}
C_{k}^\pm}{\mathrm{d} t} \mp 2f_{k}^\pm C_{k}^\pm =0, \qquad
f_{k}^\pm = f(\lambda _{k}^\pm).
\end{equation}

In fact we will analyze the various possible types of one-soliton
solutions; in our case they are determined by one {\em pair} of discrete
eigenvalues $\lambda ^\pm \in \bbbc_\pm $ and one {\em pair} of
norming constants $C^\pm $. Thus for the generic NLEE (\ref{eq:NLEE0})
we get just one type of one-soliton solutions. In order to derive its
explicit form we shall use the dressing Zakharov-Shabat method \cite{ZS-dress}.
In our case the dressing factor $u(x,t,\lambda)$ is given by a $2\times 2$
matrix of the form (\ref{dresfac}) where $P$ is a projector of rank 1 (see formula
(\ref{dresfac_1})). Then the following relations hold
\begin{eqnarray}\label{eq:P1-lr}
P|n(x,t)\rangle  &=& |n(x,t)\rangle, \qquad |n(x,t)\rangle =
\left( \begin{array}{c}n^{1}(x,t)\\ n^{2}(x,t)\end{array}
\right), \\
\langle m(x,t)|P(x,t) &=&\langle m(x,t)|, \qquad \langle m(x,t)| =
\left(m^{1}(x,t), m^{2}(x,t) \right).
\end{eqnarray}
The transmission coefficients are transformed by the dressing procedure as follows
\begin{equation}\label{eq:u-P2}
a^{+}(\lambda) = c(\lambda ) a_0^+(\lambda ), \qquad
a(\lambda) =  \frac{a_0^+(\lambda )}{c(\lambda )},
\end{equation}
The $sl(2)$ analog of formula (\ref{q0_to_q}) reads
\begin{equation}\label{eq:q'_0}
q(x,t)-q_0(x,t) = -(\lambda^+-\lambda^-)[\sigma _3,P(x,t)].
\end{equation}
By applying the above formulae to properly choosen constant vectors $|n_{0}\rangle  $
and $\langle m_{0}| $ we can construct the eigenvectors of $P(x,t) $
and as a result, obtain $P(x,t) $ explicitly. It then remains only to insert
it into eq. (\ref{eq:q'_0}) in order to obtain the corresponding potential
$q(x,t) $ explicitly.  It can be proved that the spectrum of $L(\lambda )$
will differ from the spectrum of $L_0(\lambda ) $ only by an additional pair of
discrete eigenvalues located at $\lambda^\pm \in \bbbc_\pm$.

A pure soliton solution is obtained by assuming $q_0(x,t)=0 $; as a
result we have:
\begin{eqnarray}\label{eq:1s-00}
|n(x,t)\rangle &=& e^{(-i(x\lambda^+ + f^+t)
\sigma _3} |n_{0}\rangle, \qquad \langle m(x,t)| = \langle
m_{0}|e^{i(x\lambda
^- +f^-t)\sigma _3} ,\nonumber\\
\label{eq:1s-01} P(x,t) &=& \frac{1}{ 2\cosh \Phi_0 (x,t) }
\left(\begin{array}{cc} e^{\Phi_0 (x,t)} & \kappa _2
e^{-i\Phi (x,t)} \\ \frac{1}{\kappa _2}
e^{i\Phi (x,t)} & e^{-\Phi_0 (x,t)} \end{array}\right) \\
\label{eq:1s-02} \Phi_0 (x,t) &=& -i(\lambda^+ -\lambda^-) x
-i(f^+-f^-)t -\ln \kappa _1, \nonumber\\
\Phi (x,t) &=& (\lambda^+ +\lambda ^-) x
+(f^+ + f^-)t , \nonumber
\end{eqnarray}
where $f^\pm  $ and the constants $\kappa_1 $ and $\kappa _2 $ are given
by:
\begin{equation}\label{eq:k-0-}
f^{\pm} = f(\lambda^\pm), \qquad
\kappa _1 =  \sqrt{\frac{n_{0}^1 m_{0}^1}{n_{0}^2 m_{0}^2}}, \qquad
\kappa _2 = \sqrt{\frac{n_{0}^1 m_{0}^2}{n_{0}^2 m_{0}^1}}.
\end{equation}
Then the corresponding one-soliton solution takes the form:
\begin{equation}\label{eq:0-1s}
q^{+}(x,t) = -\frac{\kappa _2 (\lambda^+ - \lambda^-)
e^{-i\Phi (x,t)}}{\cosh \Phi _0(x,t)}, \qquad
q^{-}(x,t) = \frac{\kappa _2 (\lambda^+ - \lambda^-)
e^{i\Phi (x,t)}}{\kappa _2\cosh \Phi _0(x,t)}.
\end{equation}

\begin{remark}\label{rem:1}
The eigenvalues $\lambda^\pm $ are two independent complex
numbers, therefore in the denominator in eq. (\ref{eq:1s-01}) we
get $\cosh $ of complex argument. This function vanishes whenever
its argument becomes equal to $i(\pi/2 +p\pi) $ for some integer
$p $ and the generic solitons of the (\ref{eq:NLEE0}) may have
singularities for finite $x $ and $t $.
\end{remark}

One way to avoid these singularities is to impose on the Zakharov-Shabat
system an involution, i.e. if we constrain the potential $q_0(x,t) $ by:
\begin{equation}\label{eq:invol}
q(x,t) = q^\dag (x,t), \qquad \mbox{i.e.\;} \qquad q^+ = (q^-)^*
=u(x,t).
\end{equation}

Such constraint reduces the generic systems (\ref{eq:NLEE0}) to NLEE for
the single function $u(x,t) $; the second equation of the system becomes
consequence of the first one. As a result eq. (\ref{eq:1.35}) becomes the
NLS eq.:
\begin{equation}\label{eq:nls}
i u_t + u_{xx} + 2|u|^2u(x,t)=0
\end{equation}
while eq. (\ref{eq:1.37}) goes into the MKdV-type equation:
\begin{equation}\label{eq:MkDv0}
u_{t} + u_{xxx} + 6|u(x,t)|^2u_{x} = 0.
\end{equation}
This involution imposes constraints on all the scattering data; in
particular we have:
\begin{equation}\label{eq:scat-in}
a^+(\lambda ) = (a^-(\lambda ^*))^*, \qquad
b^+(\lambda ) = (b^-(\lambda ^*))^*,
\end{equation}
>From the first relation above we find that the zeroes of the functions
$a^\pm(\lambda ) $ which are the eigenvalues of $L_0(\lambda ) $ must
satisfy:
\begin{equation}\label{eq:invo-sd}
\lambda^+= (\lambda ^-)^* =\mu + i\nu,
\qquad C^+=(C^-)^*, \qquad P(x,t) =
P^\dag(x,t).
\end{equation}
So now the one-soliton solution corresponds to a pair of eigenvalues
which must be mutually conjugated pairs.

As a result we find that the expression for $P(x,t) $ and the one for
the one-soliton solution  simplifies to
\begin{eqnarray}\label{eq:1s-00i}
P(x,t) &=& \frac{1}{2\cosh \Phi _{00}(x,t) } \left(\begin{array}{cc}
e^{\Phi _{00}(x,t)} & e^{-i\Phi _{01}(x,t)} \\
e^{i\Phi _{01}(x,t)} & e^{-\Phi _{00}(x,t)} \end{array}\right) \\
\label{eq:1s-02i}
\Phi _{00}(x,t) &=& 2\nu x + 2ht -\ln \left| \frac{n_{0}^1}{
n_{0}^2 }\right| ,\nonumber\\
\Phi _{01}(x,t) &=& 2\mu x + 2g t -\arg n_{0}^1 + \arg
n_{0}^2  ,\nonumber
\end{eqnarray}
where
\begin{equation}\label{eq:fgh}
\lambda^{\pm} = \mu \pm i\nu, \qquad
f^{\pm} = g \pm i h.
\end{equation}
Now both functions $\Phi _{00}(x,t) $ and $\Phi _{01}(x,t)$ become real
valued.  The denominator now becomes $\cosh $ of real argument, so this
soliton solution is regular function for all $x $ and $t $.

One can impose on $q_0(x,t) $ a different involution
\begin{equation}\label{eq:invol'}
q(x,t) = -q^\dag (x,t), \qquad \mbox{i.e.\;} \qquad q^+ = -(q^-)^*
=u(x,t).
\end{equation}
However it is well known that under this involution the Zakharov Shabat
system $L(\lambda ) $ becomes equivalent to an eigenvalue problem:
\begin{equation}\label{eq:eigL0}
\mathcal{L} \psi (x,t,\lambda ) \equiv i\sigma _3 \partial_x\psi
 + \sigma _3q(x,t)\psi (x,t,\lambda ) = \lambda \psi
(x,t,\lambda ),
\end{equation}
where the operator $\mathcal{L} $ is a self-adjoint one, so its spectrum
must be on the real $\lambda  $-axis. But the continuous spectrum
of $\mathcal{L} $ fills up the whole real $\lambda $-axis, which leaves no
room for solitons.

Finally, the Zakharov-Shabat system can be restricted by a third
involution, e.g.
\begin{equation}\label{eq:invol''}
q(x,t) =-q^T(x,t), \qquad \mbox{i.e.\,} \qquad q^+=-q^-=-iw_x.
\end{equation}
Such involution is compatible only with those NLEE whose dispersion
law is odd function $f(\lambda )=-f(-\lambda ) $. Therefore it can not
be applied to the NLS eq.; applied to the MKdV eq. it gives:
\begin{equation}\label{eq:MkDv0'}
w_{xt} + w_{xxxx} + 6(w_x(x,t))^2w_{xx} = 0,
\end{equation}
which can be integrated ones with the result $v=w_x $:
\begin{equation}\label{eq:MkDv0''}
v_{t} + v_{xxx} + 6(v(x,t))^2v_{x} = 0,
\end{equation}
i.e. we get the MKdV eq. for the real-valued function $v(x,t) $.
It is well known also that the NLEE with dispersion law $f(\lambda
)=(2\lambda )^{-1} $ can be explicitly derived under this reduction and
comes out to be the famous sine-Gordon eq. \cite{AKNS0}:
\begin{equation}\label{eq:sg}
w_{xt} + \sin (2w(x,t))=0.
\end{equation}

This second involution can be imposed together with the one in
(\ref{eq:invol}). The restrictions that it imposes on the scattering data
are as follows:
\begin{equation}\label{eq:scat-in''}
a^+(\lambda ) = (a^-(\lambda ^*))^*, \qquad
a^+(\lambda )=(a^-(-\lambda)),
\end{equation}
Now if $\lambda^+ $ is an eigenvalue of $L(\lambda ) $ then
$(\lambda^+)^* $, $-\lambda^+ $ and $-(\lambda^+)^*
$ must also be eigenvalues. This means that we can have two configurations
of eigenvalues:
\begin{enumerate}

\item pairs of purely imaginary eigenvalues:
\begin{equation}\label{eq:im-lam}
\lambda^+ = i\nu \equiv -(\lambda^+)^*,
\qquad \lambda^- = -i\nu  \equiv -(\lambda^-)^*,
\end{equation}

\item quadruplets of complex eigenvalues:
\begin{eqnarray}\label{eq:im-lam'}
\lambda^+ &=& \mu + i\nu \qquad -(\lambda
^+)^*
= -\mu  +i\nu, \nonumber\\
\lambda ^- &=& \mu -i\nu ,  \qquad -(\lambda
^-)^* =-\mu -i\nu.
\end{eqnarray}

\end{enumerate}

Thus we conclude, that the sine-Gordon and MKdV equations allow {\em two
types} of solitons: type 1 with purely imaginary pairs of eigenvalues and
type 2 each corresponding to a quadruplet of eigenvalues. Type 1 solitons
are known also as topological solitons, or kinks (for details see \cite{FaTa}). They
are parametrized by two real parameters: $\nu $ and $|C^+| $
so they have just one degree of freedom corresponding to the uniform
motion.

Type 2 solitons are known as the breathers and are parametrized by 4 real
parameters:  $\mu$ and $\nu $ and the real and imaginary
parts of $C^+ $.  Therefore they have two degrees of freedom: one
corresponds to the uniform motion and the second one describes the
internal degree of freedom responsible for the `breathing'.

The purpose of presenting the above well-known facts in the above manner,
was simply to make it clear that the structure, as well as the number of
related parameters which determine what different types of solitons can
exist, depend strongly on the type of, and the number of, different
involutions that can be imposed on the system.

\section{$N$-wave system related to $sl(3)$}

In this subsection we are going to consider generic $N$-wave system
related to $sl(3)$ along with its reductions. The 1-soliton solutions
to this system are about to be derived as well. It proves to be
convenient not to use a standard matrix notation but a notation which
exploites the root structure of $sl(3)$, namely $Q_{kn}$ $k, n=1,2$
stands for the component of $Q$ associated with the root
$\alpha=k\alpha_1+n\alpha_2$ expanded over the the simple roots
$\alpha_1=e_1-e_2$ and $\alpha_2=e_2-e_3$. Taking into account that
convention the $sl(3)$-$N$-wave system  consists of 6 equations of the form
\begin{equation}
\begin{split}
i(J_1-J_2)Q_{10,t}-i(I_1-I_2)Q_{10,x}+3kQ_{11}Q_{\overline{01}}=0,\\
i(2J_1+J_2)Q_{11,t}-i(2I_1+I_2)Q_{11,x}-3kQ_{10}Q_{01}=0,\\
i(2J_2+J_1)Q_{01,t}-i(2I_2+I_1)Q_{01,x}+3kQ_{\overline{10}}Q_{11}=0,\\
\end{split}\end{equation}
where $k=J_1I_2-I_1J_2$ is an arbitrary constant. The rest of the 6-wave
equations can be obtained by using the following tranformation:
$Q_{kn}\leftrightarrow Q_{\overline{kn}}$, where $Q_{\overline{kn}}=Q_{-k,-n}$.
This system can be solved via a dressing procedure with the dressing factor
(\ref{dresfac}). The 1-soliton solution obtained that way is given
by the following expressions
\begin{eqnarray}
Q_{10}(z)&=&\frac{\lambda^{-}-\lambda^{+}}{\langle m|n\rangle}
e^{-i(\lambda^{+}z_1-\lambda^{-}z_2)}n_0^1 m_0^2,\nonumber\\
Q_{11}(z)&=&\frac{\lambda^{-}-\lambda^{+}}{\langle m|n\rangle}
e^{-i(\lambda^{+}z_1+\lambda^{-}(z_1+z_2))}n_0^1 m_0^3,\label{gen_1sol}\\
Q_{01}(z)&=&\frac{\lambda^{-}-\lambda^{+}}{\langle m|n\rangle}
e^{-i(\lambda^{+}z_2+\lambda^{-}(z_1+z_2))}n_0^2 m_0^3,\nonumber
\end{eqnarray}
where
\[\langle m|n\rangle =\sum^3_{j=1}e^{-i(\lambda^+ -\lambda^-)z_j}n^j_0 m^j_0,
\qquad z_{\sigma}=J_{\sigma}x+I_{\sigma}t,\qquad \sigma=1,2.\]
The other three fields can be derived from these by executing the following
change of variables
\[Q_{kn}\leftrightarrow Q_{\overline{kn}},\qquad e^{i\lambda^{+}z_j}\leftrightarrow
e^{-i\lambda^{-}z_j},\qquad n^j_0\leftrightarrow m^j_0.\]

Impose a $\bbbz_2$ reduction of the type
\begin{equation}
K_1U^{\dag}(x,\lambda^{\ast})K^{-1}_1=U(x,\lambda),
\quad\Rightarrow\, K_1J^{\ast}K^{-1}_1=J,\quad
K_1Q^{\dag}K^{-1}_1=-Q\label{sym_pot1},
\end{equation}
where $K_1=\diag(\epsilon_1,\epsilon_2,\epsilon_3)$ is an element
of the Cartan subgroup $H\subset SL(3)$ which represents an action of $\bbbz_2$.
This results in reducing the number of independent fields since we have
\[Q_{\overline{10}}=-\epsilon_1\epsilon_2 Q^{\ast}_{10},
\qquad Q_{\overline{11}}=-\epsilon_1\epsilon_3 Q^{\ast}_{11},
\qquad Q_{\overline{01}}=-\epsilon_2\epsilon_3 Q^{\ast}_{01}\]
and therefore the number of equations from 6 to 3 as follows
\begin{equation}
\begin{split}
i(J_1-J_2)Q_{10,t}-i(I_1-I_2)Q_{10,x}-3k\epsilon_2\epsilon_3 Q_{11}Q^{\ast}_{01}=0,\\
i(2J_1+J_2)Q_{11,t}-i(2I_1+I_2)Q_{11,x}-3kQ_{10}Q_{01}=0,\\
i(2J_2+J_1)Q_{01,t}-i(2I_2+I_1)Q_{01,x}-3k\epsilon_1\epsilon_2 Q^{\ast}_{10}Q_{11}=0.
\end{split}\end{equation}
The discrete eigenvalues of $\bbbz_2$-reduced operator $L$ are complex conjugated, i.e.
$\lambda^{-}=(\lambda^{+})^{\ast}=\mu-i\nu$ and the polarization vectors are interrelated via
$|n\rangle =K_1|m\rangle^{\ast}$. The 1-soiton solution in this case is
\begin{equation}
\begin{split}
Q_{10}(z) = -\frac{2i\nu e^{\nu(z_1+z_2)}}{\langle n|^{\ast}K_1|n\rangle}
e^{-i\mu(z_1-z_2)}n_0^1 \epsilon_2(n_0^2)^{\ast},\\
Q_{11}(z) =- \frac{2i\nu e^{-\nu z_2}}{\langle n|^{\ast}K_1|n\rangle}
e^{-i\mu(2z_1+z_2)}n_0^1 \epsilon_3(n_0^3)^{\ast},\\
Q_{01} (z)= -\frac{2i\nu e^{-\nu z_1}}{\langle n|^{\ast}K_1|n\rangle}
e^{-i\mu(2z_2+z_1)}n_0^2 \epsilon_3(n_0^3)^{\ast}.
\end{split}\end{equation}

\begin{remark}
In general, the denominator of the expressions for the 1-soliton solution
(\ref{gen_1sol}) can possess zeros for some $x$ and $t$, i.e. we have
singular solutions (exploding solitons). By imposing a certain reduction
this effect can be annihilated. As it is seen from the solutions to the
$\bbbz_2$-reduced problem with $K_1=\openone$ we obtain a sum real exponents
multiplied by some positive factors which do not vanish on the real axis.
\end{remark}

By imposing another $\bbbz_2$ reduction on the potential $U(x,\lambda)$, namely
\begin{equation}
K_2U^T(x,-\lambda)K^{-1}_2=-U(x,\lambda),\qquad
K_2J^TK^{-1}_2=J,\qquad K_2Q^TK^{-1}_2=Q.\label{sym_pot2}
\end{equation}
where $K_2\in H$ safisfies $[K_1,K_2]=0$ we obtain a $\bbbz_2\times\bbbz_2$-reduced
$sl(3)$ $N$-wave system. As a consequence we have a pair of purely imaginery eigenvalues
$\lambda^{\pm}=\pm i\nu$. Choosing $K_1=K_2=\openone$ we see that the three independent
fields $Q_{10}(x,t)$, $Q_{01}(x,t)$ and $Q_{11}(x,t)$ are purely imaginary while the polarization vector
is real, $|n\rangle^{\ast}=|n\rangle$. After introducing new variables
\[Q_{10}(x,t)=i\mathbf{q}_{10}(x,t),\qquad Q_{01}(x,t)=i\mathbf{q}_{01}(x,t),
\qquad Q_{11}(x,t)=i\mathbf{q}_{11}(x,t),\]
we derive a real 3-wave system for 3 real valued fields
\begin{equation}
\begin{split}
(J_1-J_2)\mathbf{q}_{10,t}-(I_1-I_2)\mathbf{q}_{10,x}
+3k\mathbf{q}_{11}\mathbf{q}_{01}=0,\\
(2J_1+J_2)\mathbf{q}_{11,t}-(2I_1+I_2)\mathbf{q}_{11,x}
-3k\mathbf{q}_{10}\mathbf{q}_{01}=0,\\
(2J_2+J_1)\mathbf{q}_{01,t}-(2I_2+I_1)\mathbf{q}_{01,x}
+3k\mathbf{q}_{10}\mathbf{q}_{11}=0.
\end{split}\end{equation}
Since the dressing factor must satisfy the conditions
\begin{eqnarray}
\left(u^{\dag}(x,\lambda^{\ast})\right)^{-1} & = & u(x,\lambda),\\
\left(u^T(x,-\lambda)\right)^{-1} & = & u(x,\lambda),
\end{eqnarray}
the projector $P$ is real valued. In this case the discrete eigenvalues
are purely imaginary, i.e. $\lambda^{\pm}=\pm i\nu$. The 1-soliton solution is
\[\mathbf{q}^{1s}_{kl}(x)=-2\nu P_{kl}(x),\qquad
 P=\frac{|n\rangle\langle n|}{\langle n|n\rangle},\qquad k\neq l.\]
Taking into account that $|n\rangle=e^{\nu Jx}|n_0\rangle$ we derive explicitly
the following result
\begin{equation}
\begin{split}
\mathbf{q}_{10}(x,t)=-\frac{2\nu e^{\nu(z_1+z_2)}n^1_0 n^2_0}{e^{2\nu z_1}(n^{1}_{0})^2
+e^{2\nu z_2 }(n^{2}_{0})^2+e^{-2\nu(z_1+z_2)}(n^{3}_{0})^2},\\
\mathbf{q}_{11}(x,t)=-\frac{2\nu e^{-\nu z_2}n^1_0 n^3_0}{e^{2\nu z_1}(n^{1}_{0})^2
+e^{2\nu z_2 }(n^{2}_{0})^2+e^{-2\nu(z_1+z_2)}(n^{3}_{0})^2},\\
\mathbf{q}_{01}(x,t)=-\frac{2\nu e^{-\nu z_1}n^2_0n^3_0}{e^{2\nu z_1}(n^{1}_{0})^2
+e^{2\nu z_2 }(n^{2}_{0})^2+e^{-2\nu(z_1+z_2)}(n^{3}_{0})^2}.
\end{split}\label{sldoublon}\end{equation}

In the $\bbbz_2\times\bbbz_2$ case there exists another type of soliton solutions
--- these obtained by using a dressing factor of the form
\begin{eqnarray}
u(x,\lambda)&=&\openone+(c(\lambda)-1)P(x)+\left(\frac{1}{c(-\lambda)}-1\right)
K_2P^T(x)K^{-1}_2\nonumber\\&=&\openone
+\frac{\lambda^{-}-\lambda^{+}}{\lambda-\lambda^{-}}P(x)
+\frac{\lambda^{-}-\lambda^{+}}{\lambda+\lambda^{+}}K_2P^T(x)K^{-1}_2,
\qquad P(x)=\frac{| n(x)\rangle\langle m(x)|}{\langle m(x)| n(x)\rangle}.\label{dressfac2}
\end{eqnarray}
These solutions are associated with 4 discrete eigenvalues of the scattering operator $L$:
$\pm\lambda^{\pm}$. In this sense they may be called quadruplet solitons
unlike the solutions (\ref{sldoublon}) which being associated with 2 eigenvalues
$\pm i\nu$ represent doublet solitons. The vectors
$|n(x)\rangle$ and $|m(x)\rangle$ depend on the fundamental analytic solutions
$\chi^{\pm}_0(x,\lambda)$ in the same manner as it is shown in (\ref{vec2}). The
dressing factor (\ref{dressfac2}) is $\bbbz_2\times\bbbz_2$ invariant if the conditions
hold true
\begin{eqnarray}
K_1\left(u^{\dag}(x,\lambda^{\ast})\right)^{-1}K^{-1}_1=u(x,\lambda),\\
K_2\left(u^{T}(x,-\lambda)\right)^{-1}K^{-1}_2=u(x,\lambda),
\end{eqnarray}
provided that $\lambda^{+}=(\lambda^{-})^{\ast}=\mu+i\nu$ and $K_1P^{\dag}K^{-1}_1=P$ are
satisfied. Moreover, we assume that the projectors $P$ and $K_2P^TK^{-1}_2$ are
pairwise orthogonal, i.e. $PK_2P^T=0$ is valid. The resrictions on the projectors
give rise to some algebraic relations on the polarization vectors, namely
\[|m_0\rangle=K_1|n_0\rangle^{\ast},\qquad \langle n_0|K_2|n_0\rangle= 0.\]

To find the 1-soliton solution we take the limit $\lambda\to\infty$ in equation
(\ref{dress_pde}) and put $q_0\equiv 0$. Thus we obtain the following formula
\begin{equation}
Q^{\mathrm{1s}}_{jk}=(\lambda^{-}-\lambda^{+})(P+K_2P^TK^{-1}_2)_{jk},
\qquad j\neq k.
\end{equation}
Let $K_1=K_2=\openone$. Then $Q^{\ast}=-Q$ and using the above notation we have for
the 1-soliton solution
\begin{eqnarray}
\mathbf{q}_{10}&=&-\frac{4\nu|n_{0}^{1}n_{0}^{2}|}{\langle n|n\rangle^{\ast}}e^{\nu(z_1+z_2)}
\cos[\mu(z_1-z_2)-\delta_1+\delta_2],\nonumber\\
\mathbf{q}_{11}&=&-\frac{4\nu|n_{0}^{1}n_{0}^{3}|}{\langle n|n\rangle^{\ast}}
e^{-\nu z_2}\cos[\mu(2z_1+z_2)-\delta_1+\delta_3],\\
\mathbf{q}_{01}&=&-\frac{4\nu|n_{0}^{2}n_{0}^{3}|}{\langle n|n\rangle^{\ast}}
e^{-\nu z_1}\cos[\mu(2z_2+z_1)-\delta_2+\delta_3],\nonumber\\
\langle n|n\rangle^{\ast}&=&e^{2\nu z_1}|n_{0}^{1}|^2+e^{2\nu z_2}|n_{0}^{2}|^2
+e^{-2\nu(z_1+z_2)}|n_{0}^{1}|^3,\qquad n_{0}^{j}=|n_{0}^{j}|e^{i\delta_j}.\nonumber
\end{eqnarray}

\section{Generalized Zakharov-Shabat system and $sl(n) $
solitons}\label{sec:3}

For the sake of simplicity and clarity below, most of our discussions
will be restricted to the case $n=5 $; however they also could easily be
reformulated for any other chosen value of $n $. The corresponding Lax
operator $L(\lambda ) $ which is a particular case of eq. (\ref{eq:L-})
with
\begin{eqnarray}
L\equiv i\partial_x+U(x,t,\lambda ) &=& i\partial_x+[J,Q(x,t)]-\lambda J, \\
J=\diag(J_1 , J_2, J_3, J_4, J_5),& & Q(x,t) = \left( \begin{array}{ccccc} 0 &
Q_{12} & Q_{13} & Q_{14} & Q_{15}\\ Q_{21} & 0  & Q_{23} & Q_{24} &
Q_{25}\\ Q_{31} & Q_{32} & 0 & Q_{34} & Q_{35}\\ Q_{41} & Q_{42} & Q_{43}
 & 0 & Q_{45}\\ Q_{51} & Q_{52} & Q_{53} & Q_{54} &0 \end{array}\right),
\label{eq:U-5}\end{eqnarray}
Furthermore, for definiteness we will assume that
\begin{equation}\label{eq:J-type}
\tr J=0 , \qquad J_1>J_2>J_3 >0,  \qquad 0>J_4>J_5.
\end{equation}

The $M $-operator in the Lax representation for the $N $-wave
equation (\ref{eq:N-w0}) is given by:
\begin{equation}\label{eq:M-0}
M\psi (x,t,\lambda )\equiv  i \partial_t\psi + ([I,
Q(x,t)] -\lambda I) \psi (x,t,\lambda ) = -\lambda \psi
(x,t,\lambda )I,
\end{equation}
where $I=\diag(I_1,\hdots,I_5)$ is a traceless matrix.

As we discussed in the section Preliminaries the 1-soliton solution
can be derived by using formula (\ref{q0_to_q})
\begin{equation}\label{eq:q'}
q(x) = \lim_{\lambda \to\infty } \lambda \left( J- u(x,\lambda )
Ju^{-1}(x,\lambda )\right)= -(\lambda^+-\lambda^-)[J,P(x)].
\end{equation}
where the projector $P$ is of the form
\begin{equation}\label{eq:ZS-1}
P(x) = \frac{|n(x)\rangle  \langle m(x)|}{\langle
m(x)| n(x)\rangle   } ,\quad
|n(x)\rangle  = \chi ^+_0(x,\lambda^+) |n_{0} \rangle
, \quad\langle m(x)| = \langle m_{0}|
\hat{\chi }^-_0(x,\lambda^-).
\end{equation}
The polarization vectors $|n_0\rangle$ and $\langle m_0|$ are constant 5-vectors.
The 1-soliton solution is parametrized by:
\begin{enumerate}

\item the discrete  eigenvalues $\lambda^\pm =\mu^{\pm}\pm i\nu^{\pm}
$; $\mu^{\pm}$ determine the soliton velocity, $\nu ^\pm $
determine the amplitude.

\item the `polarization' vectors. $|n_{0}\rangle , \langle
m_{0}| $ parametrize the internal degrees of freedom of the
soliton. Note that $P(x) $ is invariant under the scaling of each of
these vectors.  Generically each `polarization' has 5 components, one of
which can be fixed, say to 1.  So each `polarization' is determined by
4 independent complex parameters.

\end{enumerate}

We have several options that will lead to different types of solitons:

-- 1) generic case when all components of $|n_{0}\rangle  $
are non-vanishing;

-- 2) several special subcases when one (or several) of these components
vanish. The corresponding solitons will have different structures and
properties.

For the generic choice of $|n_{0}\rangle  $ one finds:
\begin{equation}\label{eq:P1-pm}
\lim_{x\to\pm\infty } P(x,t) = P_{\pm}, \qquad
P_{+} = E_{11}, \qquad P_{-}=E_{nn},
\end{equation}
where the matrix $E_{kj} $ has only one non-vanishing matrix element equal
to 1 at position $k,j $, i.e. $(E_{kj})_{mp}=\delta _{km}\delta _{jp} $.
Therefore both the limiting values $u_\pm(\lambda ) $ and their inverse
$\hat{u}_\pm(\lambda ) $ are diagonal matrices:
\begin{eqnarray}\label{eq:u-hu}
u_+(\lambda ) &=& \diag (c(\lambda ), 1,1,\dots ,1), \qquad
u_-(\lambda ) = \diag (1,1,\dots ,1, c(\lambda ) ).
\end{eqnarray}
>From eqs. (\ref{T_dress}) for $n=5 $ we have
\begin{eqnarray}\label{eq:36a}
T_{1j}(\lambda ) &=& c(\lambda ) (T_0)_{1j}(\lambda ), \qquad j=1,2,3,4;
\nonumber\\
T_{j5}(\lambda ) &=&  (T_0)_{j5}(\lambda ) /c(\lambda ),  \qquad
j=2,3,4,5;\\
T_{ij}(\lambda ) &=&  (T_0)_{ij}(\lambda ), \qquad \mbox{for all other
values of $i,j $.}
\end{eqnarray}

This relation allows us to derive the interrelations between the
Gauss factors of $T_0(\lambda ) $ and $T(\lambda ) $. In particular we find for the
principal minors of $T(\lambda ) $:
\begin{equation}\label{eq:m_k+}
m_k^{+} (\lambda ) = c(\lambda )m_{0,k}^+(\lambda ) ,\qquad
m_k^{-}(\lambda ) =m_{0,k}^-(\lambda )/ c(\lambda ) ,
\end{equation}
where $m_{k}^{+}(\lambda ) $ (resp. $m_{k}^{-}(\lambda ) $)
are the upper (resp. lower) principal minors of $T(\lambda ) $. Since
$\chi_0 ^\pm(x,t,\lambda ) $ are regular solutions of the RHP then
$m_{0,k}^\pm(\lambda ) $ have no zeroes at all, but eq.  (\ref{eq:m_k+})
means {\em all} $m_{k}^{\pm}(\lambda ) $ have a simple zero at
$\lambda =\lambda^\pm $.

The generic one-soliton solution then is obtained by taking $\chi
^\pm(x,t,\lambda ) = e^{-i\lambda (Jx+It)} $. As a result we get:
\begin{eqnarray}\label{eq:G-1s-P}
(P(x,t))_{ks} &=& \frac{1}{k(x,t)}  n_{0}^{k} m_{0}^{s}
e^{-i(\lambda^+ z_k-\lambda^-z_s)}  ,\\
k(x,t) &=&\sum_{p=1}^{n} n_{0,1}^{p} m_{0,1}^{p}
e^{-i(\lambda^{+} -\lambda^{-})z_p(x,t) },\\
\label{eq:G-1s-P'}
z_k(x,t) &=& J_kx +I_kt, \qquad
q_{ks}^{\rm 1s} = -(\lambda^+ -\lambda^-)
(P(x,t))_{ks},
\end{eqnarray}
i.e. in all channels we have non-trivial waves. The number of internal
degrees of freedom is $2(n-1)=8 $. Note that the denominator $k(x,t) $ is
a linear combination of exponentials with complex arguments, so it could
vanish for certain values of $x $, $t $. Thus the generic soliton
(\ref{eq:G-1s-P}) in this case is a singular solution.

Next we impose on $U(x,t,\lambda ) $ the involution:
\begin{equation}\label{eq:B0}
K U^\dag (x,t,\lambda ^*) K^{-1} =U(x,t,\lambda ), \qquad K = \diag
(\epsilon _1, \dots, \epsilon _n),
\end{equation}
with $\epsilon _j=\pm 1 $. More specifically this means that:
\begin{equation}\label{eq:B01}
Kq^\dag (x,t) K^{-1} = q(x,t), \qquad K u^\dag (x,t,\lambda ^*)
K^{-1} = u^{-1}(x,t,\lambda ),
\end{equation}
and
\begin{equation}\label{eq:inv-2}
\lambda^+ = (\lambda^-)^* = \mu+i\nu, \qquad  \langle
m_{0}| = (K |n_{0} \rangle )^\dag .
\end{equation}
Thus only $|n_{0}\rangle  $ is independent.

Then the one-soliton solution simplifies to:
\begin{eqnarray}\label{eq:q1s-red}
q_{ks}^{\rm 1s}(x,t) &=& -\frac{2i\nu(J_k-J_s)}{ k_{\rm
red}(x,t) } \epsilon _s n_{0}^{k} (n_{0}^{s})^* e^{\nu(z_k+z_s)} e^{-i\mu
(z_k-z_s)} ,\\
k_{\rm red}(x,t) &=& \sum_{p=1}^{n} \epsilon _p |n_{0}^{p}|^2
e^{2\nu z_p(x,t)}.
\end{eqnarray}
The number of internal degrees of freedom now is $n-1=4 $. If one or more
of $\epsilon_j $ are different, then this reduced soliton may still have
singularities. The singularities are absent only if all $\epsilon _j $ are
equal.

\subsection{Non-generic  $sl(2)$  solitons.}\label{ssec:sl2}

>From now on we assume that the reduction (\ref{eq:B0}) with $\epsilon _p=1
$ holds. Here $|n_{0,1}\rangle  $ has only two non-vanishing components. We
consider here three examples with $n=5 $ and three different choices for
the polarization vectors:
\begin{equation}\label{eq:14.1}
\mbox{a)}\quad |n_{0,1}\rangle = \left( \begin{array}{c} n_{0}^{1}
\\ 0 \\ 0 \\ 0 \\ n_{0}^{5} \end{array} \right); \qquad
\mbox{b)}\quad |n_{0}\rangle = \left( \begin{array}{c} 0 \\
n_{0}^{2} \\ 0 \\ n_{0}^{4} \\ 0 \end{array} \right); \qquad
\mbox{c)}\quad |n_{0}\rangle = \left( \begin{array}{c} n_{0}^{1}
\\ n_{0}^{2} \\ 0 \\ 0 \\ 0 \end{array} \right) .
\end{equation}
In all these cases the corresponding one-soliton solutions $q(x,t) $ are
given by similar analytic expressions, each
having only two non-vanishing matrix elements:
\begin{eqnarray}\label{eq:14.3}
q_{jk}(x,t) &=& (q_{jk}(x,t))^* \nonumber\\
&=& -\frac{i\nu(J_j-J_k) e^{i (\arg (n_{0}^{j}) - \arg
(n_{0}^{k})) } e^{-i\mu(J_j-J_k) (x+w_{jk}t)} }{
\cosh [ \nu(J_j-J_k) (x+w_{jk}t) + \ln |n_{0}^{j}| - \ln
|n_{0}^{k}|] },
\end{eqnarray}
where we remind that $w_{jk} = (I_j-I_k)/(J_j-J_k) $, $j<k $. For the case
a) we have $j=1 $, $k=5 $; in case b): $j=2 $, $k=4 $ and in case c) $j=1
$ and $k=2 $.

The $sl(2) $ soliton is very much like the NLS soliton (apart from the $t
$-dependence); the NLS soliton has only one internal degree of freedom.

The different choices for the polarization vector result in different
asymptotics for the projector $P_1(x,t) $:
\begin{eqnarray}\label{eq:14.2}
\mbox{a)}\quad \lim_{x\to\infty } P(x,t) = E_{11}, \qquad \lim_{x\to
-\infty } P(x,t) = E_{55}, \nonumber\\
\mbox{b)}\quad \lim_{x\to\infty } P(x,t) = E_{22}, \qquad \lim_{x\to
-\infty } P(x,t) = E_{44}, \nonumber\\
\mbox{c)}\quad \lim_{x\to\infty } P(x,t) = E_{11}, \qquad \lim_{x\to
-\infty } P(x,t) = E_{22},
\end{eqnarray}

In case a) the results for the limits of $P(x,t) $ and for
$u_\pm(\lambda ) $ are the same as for the generic case, see eqs.
(\ref{eq:P1-pm}), (\ref{eq:u-hu}). As a consequence, such $sl(2) $
solitons requires the vanishing of {\it all} Evans functions
$m_k^{\pm}(\lambda ) $ for $\lambda =\lambda^\pm $, see eq.
(\ref{eq:m_k+}).

In case b) from eq. (\ref{T_dress}) and from the appendix we get
that such $sl(2) $ soliton provides for the vanishing of
$m_2^\pm(\lambda ) $ and $m_3^\pm(\lambda ) $:
\begin{eqnarray}\label{eq:15.5}
m_2^{+}(\lambda ) &=& c(\lambda ) m_{0,2}^+(\lambda ), \qquad
m_3^{+}(\lambda ) = c(\lambda ) m_{0,3}^+(\lambda ), \nonumber\\
m_2^{-}(\lambda ) &=&  m_{0,2}^-(\lambda )/c(\lambda ) , \qquad
m_3^{-}(\lambda ) = m_{0,3}^-(\lambda )/c(\lambda ) ,
\end{eqnarray}
whereas $ m_1^{\pm}(\lambda ) = m_{0,1}^\pm(\lambda )  $ and
$ m_4^{\pm}(\lambda ) = m_{0,4}^\pm(\lambda )  $ remain regular
and do not have zeros at $\lambda =\lambda^\pm $.

Likewise in case c)  we get  that only $m_1^{+}(\lambda ) $ and
$m_4^{-}(\lambda ) $ acquire zeroes:
\begin{eqnarray}\label{eq:15.5c}
m_1^{+}(\lambda ) &=& c(\lambda ) m_{0,1}^+(\lambda ), \qquad
m_4^{-}(\lambda ) = m_{0,4}^-(\lambda )/c(\lambda ) ,
\end{eqnarray}
and all the other Evans functions  $m_j^{+}(\lambda ) $ with
$j=2,3,4 $, and $m_p^{-}(\lambda ) $ with $p=1,2,3 $ do not have
zeroes.

\subsection{Non-generic $sl(3) $-solitons}\label{ssec:sl3}

Here $|n_{0}\rangle  $ has three non-vanishing components. We
consider three examples of such polarization vectors:
\begin{equation}\label{eq:18.1}
\mbox{a)} \quad |n_{0}\rangle = \left( \begin{array}{c} n_{0}^{1}
\\ 0 \\ n_{0}^{3} \\ 0 \\ n_{0}^{5} \end{array} \right), \qquad
\mbox{b)} \quad |n_{0}\rangle = \left( \begin{array}{c} 0\\
n_{0}^{2} \\ n_{0}^{3}  \\ n_{0}^{4} \\ 0 \end{array} \right),\qquad
\mbox{c} \quad |n_{0}\rangle = \left( \begin{array}{c} n_{0}^{1}
 \\ n_{0}^{2}\\ n_{0}^{3} \\ 0 \\ 0  \end{array} \right),
\end{equation}
Therefore the $sl(3) $-solitons have two internal degrees of freedom.

The asymptotics of the projector $P(x,t) $ read as follows:
\begin{eqnarray}\label{eq:18.2}
\mbox{a)}\quad \lim_{x\to\infty } P(x,t) = E_{11}, \qquad \lim_{x\to
-\infty } P(x,t) = E_{55}, \nonumber\\
\mbox{b)}\quad \lim_{x\to\infty } P(x,t) = E_{22}, \qquad \lim_{x\to
-\infty } P(x,t) = E_{44}, \nonumber\\
\mbox{c)}\quad \lim_{x\to\infty } P(x,t) = E_{11}, \qquad \lim_{x\to
-\infty } P(x,t) = E_{33},
\end{eqnarray}

Note that cases a) and b) in eq. (\ref{eq:18.2}) coincide with the
corresponding cases in eq. (\ref{eq:14.1}). Therefore the set of Evans
functions that acquire zeroes will be the same as for the corresponding $
sl(2) $ solitons. In case c) of eq. (\ref{eq:18.2}) we have:
\begin{eqnarray}\label{eq:19.5}
m_1^{+}(\lambda ) &=& c(\lambda ) m_{0,1}^+(\lambda ), \qquad
m_2^{+}(\lambda ) = c(\lambda ) m_{0,2}^+(\lambda ), \nonumber\\
m_4^{-}(\lambda ) &=&  m_{0,4}^-(\lambda )/c\lambda ) , \qquad
m_3^{-}(\lambda ) = m_{0,3}^-(\lambda )/c(\lambda ) ,
\end{eqnarray}
whereas the remaining Evans functions $m_j^{+}(\lambda ) $ with
$j=3,4 $, and $m_p^{-}(\lambda ) $ with $p=1,2$ remain regular.

In case a) the corresponding one-soliton solutions acquire the
form:
\begin{eqnarray}\label{eq:18.3}
&& \mbox{a)} \quad
q^{\rm 1s}(x,t) = \left( \begin{array}{ccccc}0 & 0 & q_{13} & 0 & q_{15}
\\ 0 & 0 & 0 & 0 & 0 \\ q_{13}^* & 0 & 0 & 0 & q_{35} \\ 0 & 0 & 0 & 0 & 0
\\ q_{15}^* & 0 & q_{35}^* & 0 & 0 \end{array} \right), \qquad
\mbox{b)} \quad
q^{\rm 1s}(x,t) = \left( \begin{array}{ccccc}0 & 0 &0  & 0 & 0 \\
0  & 0 &  q_{23} &  q_{24} & 0 \\
0 & q_{23}^* & 0 & q_{34} & 0 \\ 0 & q_{24}^* & q_{34}^* & 0 &  0 \\
0 & 0 &0  & 0 & 0 \end{array} \right), \nonumber\\
&& \mbox{c)} \quad  q^{\rm 1s}(x,t) = \left( \begin{array}{ccccc}
0 & q_{12} &  q_{23} & 0  & 0  \\
q_{12}^* & 0 & q_{23} & 0  & 0  \\ q_{13}^* & q_{23}^* & 0 &  0 &0 \\
0 & 0 &0  & 0 & 0 \\ 0 & 0 &0  & 0 & 0 \end{array} \right),
\end{eqnarray}
where the matrix elements $q_{ks}(x,t) $ are given by:
\begin{eqnarray}\label{eq:qks}
&& q_{ks}(x,t) = (q_{sk}(x,t))^* \nonumber\\
&=& -\frac{i\nu(J_k-J_s) \epsilon _s e^{\nu _1 (\widetilde{J}_k
+ \widetilde{J}_s) (x+\widetilde{v}_{ks}t) }
n_{0}^{k}(n_{0}^{s})^* e^{-i\mu (J_k-J_s) (x+w_{15}t)}
}{ |n_{0}^{1}|^2
e^{2\nu(\widetilde{J}_1x+\widetilde{I}_1t)} + |n_{0}^{3}|^2
e^{2\nu(\widetilde{J}_3x+\widetilde{I}_3t)} + |n_{0}^{5}|^2
e^{2\nu(\widetilde{J}_5x+\widetilde{I}_5t)} },
\end{eqnarray}
and
\begin{equation}\label{eq:J_t}
\widetilde{J}_k = J_k - (J_1+J_3+J_5)/3, \qquad
\widetilde{I}_k = I_k - (I_1+I_3+I_5)/3, \qquad \widetilde{v}_{ks} =
\frac{\widetilde{J}_k +\widetilde{J}_s }{\widetilde{I}_k +\widetilde{I}_s}.
\end{equation}
This soliton has two internal degrees of freedom and is regular.

Obviously it is by now clear how one can write down more complicated
solitons like $sl(4) $ which would be characterized by polarization
vectors of the form:
\begin{equation}\label{eq:19.1''}
\mbox{a)}\quad |n_{0,1}\rangle = \left( \begin{array}{c} n_{0}^{1}
\\ n_{0}^{2} \\ n_{0}^{3} \\ n_{0}^{4} \\ 0 \end{array} \right),
\qquad \mbox{b)}\quad |n_{0}\rangle = \left( \begin{array}{c}
n_{0}^{1} \\ n_{0}^{2} \\ n_{0}^{3} \\ 0\\ n_{0}^{5}  \end{array}
\right), \qquad \dots
\end{equation}
The $sl(4) $-solitons will have three internal degrees of freedom.

We note here that due to our choice of $J $ in (\ref{eq:J-type}), $sl(4)
$-solitons cannot give rise to generalized eigenfunctions.

\section{Eigenfunctions and eigensubspaces}\label{sec:4}

The structure of these eigensubspaces and the corresponding solitons
becomes more complicated with the growth of $n $.

In what follows we start with the generic case and split the
`polarization' vector into two parts:
\begin{eqnarray}\label{eq:n_+}
|n_{0}\rangle =|p_{0}\rangle + |d_{0}\rangle ;
\qquad
|p_{0}\rangle = \left( \begin{array}{c} n_{0}^{1} \\
n_{0}^{2} \\ n_{0}^{3} \\ 0 \\ 0 \end{array}\right), \qquad
|d_{0}\rangle = \left( \begin{array}{c} 0 \\ 0 \\ 0
\\ n_{0}^{4} \\ n_{0}^{5}  \end{array}\right).
\end{eqnarray}
and therefore
\begin{equation}\label{eq:n_+1}
|n\rangle =|p\rangle +|d\rangle ,\qquad
|p\rangle = \chi_0 ^+(x,t,\lambda^+) |p_{0}\rangle ,
\qquad
|d\rangle = \chi_0 ^+(x,t,\lambda^+) |d_{0}\rangle ,
\end{equation}
This splitting is compatible with eq. (\ref{eq:J-type}) and has
the advantage: if $\chi_0 ^+(x,t,\lambda^+)=e^{-i\lambda
^+Jx} $ then $|p\rangle $ increases exponentially for
$x\to\infty $ and decreases exponentially for $x\to -\infty  $;
$|d\rangle $ decreases exponentially for $x\to \infty  $
and increases exponentially for $x\to -\infty  $, see also the
lemma below.

What we will prove below is that one can take a special linear combination of the
columns of $\chi_0 ^+(x,t,\lambda^+ ) $ which decreases exponentially for
both $x\to\infty  $ and $x\to -\infty  $. Doing this we will use the fact
that
\begin{equation}\label{eq:fas-2}
\bchi ^+(x,t,\lambda^+ )|n_{0}\rangle \equiv
(\openone -P(x,t)) \chi ^+(x,t,\lambda^+ ) |n_{0} \rangle
= (\openone -P(x,t)) |n(x,t)\rangle  =0,
\end{equation}

\begin{lemma}\label{lem:1}
The eigenfunctions of $L $ provided by:
\begin{equation}\label{eq:fas-3}
\f^+(x,t)= \bchi ^+(x,t,\lambda^+)|p_{0}\rangle = -
\bchi ^+(x,t,\lambda^+)|d_{0}\rangle ,
\end{equation}
decrease exponentially for both $x\to\infty  $ and $x\to-\infty  $.

\end{lemma}

{\bf Proof}:
>From eq. (\ref{eq:fas-2}) and (\ref{eq:n_+}) there follows that both
expressions for $\f^+(x,t) $ coincide, so we can use each of them to ou r
advantage, see eq. (\ref{eq:fas-3}). We will use also the fact that
$\openone -P(x,t) $ is a bounded function of both $x $ and $t $.

We start with
\begin{equation}\label{eq:lim_f+}
\lim_{x\to\infty } \f^+(x,t) = \lim_{x\to\infty } \bchi^{\prime,+}
(x,t,\lambda^+)|d_{0}\rangle = (\openone -P_{+})
\lim_{x\to\infty }e^{-i\lambda^+(Jx+It)} \bbbt^-(\lambda^+) |d_{0}\rangle ,
\end{equation}
where $\bbbt^-(\lambda^+) $ is the lower triangular matrix introduced
in eq.  (\ref{fas_jost}).  If the potential is on finite support or
is reflectionless then $\bbbt^-(\lambda ) $ is rational function well
defined for $\lambda =\lambda^+ $. If the potential is generic then
$\bbbt^-(\lambda ) $ does not allow analytic continuation off the real
axis. Nevertheless $\bbbt^-(\lambda^+ ) $ can be understood as lower
triangular constant matrix (generalizing the constant $C_{0}^+ $ of the
NLS case). Being lower triangular $\bbbt^-(\lambda^+ ) $ maps
$|d_{0}\rangle $ onto $|d_{0}'\rangle =\bbbt_0^-(\lambda^+ )|d_{0}\rangle $ which is again of the form
(\ref{eq:n_+}), i.e. its first three components vanish.  Therefore
\begin{equation}\label{eq:lim_f+1}
\lim_{x\to\infty } e^{\nu ax} \f^+(x,t) = \lim_{x\to\infty } (\openone
-P_{+}) e^{\nu ax}\left( \begin{array}{c} 0 \\ 0 \\ 0 \\ e^{-i\lambda
^+(J_4x+I_4t)} n_{0}^{4,\prime} \\ e^{-i\lambda^+(J_5x+I_5t)}
n_{0}^{5,\prime} \end{array}\right) = 0,
\end{equation}
for any constant $a >0$ such that $a+J_4<0 $.

Likewize we can calculate the limit for $x\to -\infty  $:
\begin{eqnarray}\label{eq:lim_f-}
\lim_{x\to -\infty } \f^+(x,t) &=& -\lim_{x\to -\infty }
\bchi^{\prime,+}(x,t,\lambda_1^+)|p_{0}\rangle  \nonumber\\
&=& -(\openone -P_{+}) \lim_{x\to\infty }e^{-i\lambda
^+(Jx+It)} \bbbs^+(\lambda^+) |p_{0}\rangle .
\end{eqnarray}
The upper triangular matrix $\bbbs^+(\lambda^+) $ is treated
analogously as $\bbbt^-(\lambda^+) $. In the generic case it is just an
upper triangular constant matrix which maps $|p_{0}\rangle  $ onto
$|p_{0}'\rangle =\bbbs^+(\lambda^+)|p_{0}\rangle  $
whose last two components vanish. Therefore:
\begin{equation}\label{eq:lim_f-1}
\lim_{x\to-\infty } e^{\nu bx} \f^+(x,t) = \lim_{x\to -\infty
}  e^{\nu bx}(\openone -P_{-}) \left(
\begin{array}{c}e^{-i\lambda^+(J_1x+I_1t)}
n_{0}^{1,\prime} \\ e^{-i\lambda^+(J_2x+I_2t)}
n_{0}^{2,\prime} \\ e^{-i\lambda ^+(J_3x+I_3t)} n_{0}^{3 ,\prime} \\
0 \\ 0 \end{array}\right) = 0,
\end{equation}
for any constant $b <0$ such that $J_3+b>0$.

The lemma is proved.$\Box$

For the choices a) and b) of $|n_{0}\rangle  $ in eq.
(\ref{eq:14.1}) we define the square integrable  discrete eigenfunctions
using the splitting (\ref{eq:n_+}) and eq.  (\ref{eq:fas-3}).

\begin{remark}\label{rem:2}

The choice c) for $|\vec{n}_{0,1}\rangle $ does not allow
for the splitting (\ref{eq:n_+}). In this case we can introduce only
generalized discrete eigenfunctions, $\f_{\rm gen}(x,t) $, which are not
square integrable. But upon multiplying by the  exponential factor
$e^{-\nu  cx} $ with $c=(J_1+J_2)/2 $, we can obtain square integrable
functions $\f(x,t)=\f_{\rm gen}(x,t) e^{-\nu  cx} $. See also the
discussion in the next subsection.

\end{remark}

The generalized eigenfunctions come up in situations when the splitting
(\ref{eq:n_+}) is not possible, i.e. when either $|p_{0}\rangle $
or $|d_{0}\rangle $ vanish. Let us construct the generalized
eigenfunction for the polarization vector $|n_{0}\rangle  $ of case c)
in eq.  (\ref{eq:18.1}). Let $(J_1+J_2+J_3)/3 =a' $; then $J_1'=J_1-a' $,
$J_2'=J_2-a' $ and $J_3'=J_3-a' $ are such that $J_1'>J_2'>J_3' $ and
$J_1'+J_2'+J_3'=0 $.  Let us assume for definiteness that $J_1'>J_2'>0 $
and $0>J_3' $. Then we can split $|n_{0}\rangle  $ into
\begin{equation}\label{eq:n++1}
|n_{0}\rangle =|p_{0}'\rangle + |d_{0}'\rangle , \qquad
|p_{0}'\rangle = \left( \begin{array}{c} n_{0}^1 \\ n_{0}^2 \\
0 \\ 0 \\ 0 \end{array} \right),\qquad
|d_{0}'\rangle = \left( \begin{array}{c} 0 \\ 0 \\ n_{0}^3 \\ 0 \\ 0
\end{array} \right),
\end{equation}
and define
\begin{equation}\label{eq:f++}
\f^{+,\prime}(x,t) = \bchi^+(x,t,\lambda^+) |p_{0}'\rangle
 = - \bchi^+(x,t,\lambda^+) |d_{0}'\rangle,
\end{equation}
Obviously $\f^{+,\prime}(x,t) $ is an eigenfunction of the dressed
operator $L $ corresponding to the eigenvalue $\lambda _1^+ $.

Then we can prove the following lemma:
\begin{lemma}\label{lem:2}
The eigenfunction $\f^{+,\prime}(x,t) $ is such that $e^{\nu _1a'x}
\f^{+,\prime}(x,t) $ decreases exponentially for both $x\to \pm \infty $.

\end{lemma}

{\bf Proof}: The proof is similar to the one of lemma \ref{lem:1}
and we omit it.$\Box$

Since the polarization vector $|n_{0}\rangle  $ in case c) of eq.
(\ref{eq:18.1}) does not allow the splitting (\ref{eq:n_+}) the
corresponding discrete eigenfunction will not be square integrable, so it
will give rise to a generalized eigenfunction.

\section{Classification of solitons for a $N$-waves system related to
$so(5)$}\label{sec:5}

In this section we analyse how different kinds of reductions affect
the classification of the soliton solutions to a nonlinear equation.
This criterion is tightly connected with symmetries imposed on the
auxiliary linear problem (the zero curvature condition). We shall
consider in next subsections types of solitons which differ from one
another in the number of eigenvalues  associated with them: doublet
solitons associated with 2 purely imaginery eigenvalues
$\lambda^{\pm}=\pm i\nu$ and quadruplet solitons associated with 4
eigenvalues situated symmetrically with respect to the real and the
imaginery axis in $\bbbc$. This is the case when a $\bbbz_2\times\bbbz_2$
reduction is in action. Such type of reduction is compatible with the Lax
representation of a NLEE to have a dispersion law obeying $f(-\lambda)=-f(\lambda)$
(the $N$-wave equation fulfills that restriction since $f_{N-\mbox{w}}(\lambda)=-\lambda I$).

\subsection{N-wave system related to $so(5)$}

>From now on we shall focus our attention on a $N$-wave equation related
to the $so(5)$ algebra. This algebra has two simple roots
$\alpha_1=e_1-e_2$, $\alpha _2=e_2 $, and two more positive roots:
$\alpha _1+\alpha _2=e_1 $ and $ \alpha _1+2\alpha _2=
e_1+e_2=\alpha _{\rm max} $. When they come as indices, e.g. in
$Q_\alpha  $ we will replace them by sequences of two integers:
$\alpha \to kn $ if $\alpha =k\alpha _1 + n\alpha _2$. Moreover, we are
going to use the auxiliary notation $\overline{kn}=-k\alpha_1-n\alpha_2$. Thus the
$N$-wave system itself consists of 8 equations. A half of them reads
\begin{equation}\begin{split}
i(J_1-J_2)Q_{10,t}(x,t)-i(I_1-I_2)Q_{10,x}(x,t)
+k Q_{11}(x,t)Q_{\overline{01}}(x,t)=0,\\
iJ_1Q_{11,t}(x,t)-iI_1Q_{11,x}(x,t)
-k(Q_{10}Q_{01}+Q_{12}Q_{\overline{01}})(x,t)=0,\\
i(J_1+J_2)Q_{12,t}(x,t)-i(I_1+I_2)Q_{12,x}(x,t)
-kQ_{11}(x,t)Q_{01}(x,t)=0,\\
iJ_2Q_{01,t}(x,t)-iI_2Q_{01,x}(x,t)
+k(Q_{\overline{11}}Q_{12}+Q_{\overline{10}}Q_{11})(x,t)=0.\\
\end{split}\label{waves_orth_0}\end{equation}
The other 4 equations can be derived from those above by using the
formal transformation $Q_{kn}\leftrightarrow Q_{\overline{kn}}$. One is able
to integrate the system by applying the already discussed ideas --- dressing
method etc. For that purpose we make use of the dressing factor (\ref{u_orthogon}).
The 1-soliton solution reads
\begin{equation}
\begin{split}
Q_{10}(z)=\frac{\lambda^+ -\lambda^-}{\langle m|n\rangle}
\left(e^{i(\lambda^+z_2-\lambda^-z_1)}n^1_0 m^2_0+
e^{i(\lambda^-z_2-\lambda^+z_1)}n^4_0 m^5_0\right),\\
Q_{11}(z)=\frac{\lambda^+ -\lambda^-}{\langle m|n\rangle}
\left(e^{-i\lambda^-z_1}n^1_0 m^3_0-e^{-i\lambda^+z_1}n^3_0 m^5_0\right),\\
Q_{12}(z)=\frac{\lambda^+ -\lambda^-}{\langle m|n\rangle}
\left(e^{-i(\lambda^+z_2+\lambda^-z_1)}n^1_0 m^4_0+
e^{-i(\lambda^+z_1+\lambda^-z_2)}n^2_0 m^5_0\right),\\
Q_{01}(z)=\frac{\lambda^+ -\lambda^-}{\langle m|n\rangle}
\left(e^{-i\lambda^-z_2}n^2_0 m^3_0+
e^{-i\lambda^+z_2}n^3_0 m^4_0\right),\\
\langle m|n\rangle=\sum^5_{k=1}e^{i(\lambda^+ -\lambda^-)z_k}n^k_0 m^k_0,
\qquad z_{k}=J_{k}x+I_{k}t,\qquad k=1,2.
\end{split}\end{equation}
The other 4 field can be formally constructed by doing the following
transformation
\[Q_{kn}\leftrightarrow Q_{\overline{kn}},\qquad  e^{i\lambda^{+}z_k}\leftrightarrow
e^{-i\lambda^{-}z_k},\qquad n^j_0\leftrightarrow m^j_0.\]

Let us consider a $\bbbz_2$ reduction of the type
$KU^{\dag}(\lambda^*)K^{-1} =U(\lambda )$ where $K=\mbox{diag} \,
(\epsilon_1, \epsilon_2, 1, \epsilon_2, \epsilon_1)$ is an
element of the Cartan subgroup ($\epsilon_k=\pm 1 $, $k=1,2 $).
This means that $J_k=J_k^* $, $Q_\alpha  $ must satisfy:
\begin{eqnarray}\label{eq:b2.3}
Q_{\overline{10}}=-\epsilon_1\epsilon_2 Q^*_{10}, \quad Q_{\overline{01}}=-\epsilon_2Q^*_{01}, \quad
Q_{\overline{11}}=-\epsilon_1 Q^*_{11}, \quad Q_{\overline{12}}=-\epsilon_1\epsilon_2Q^*_{12}.
\end{eqnarray}
The corresponding NLEE is given by 4 equation
\begin{equation}\begin{split}
i(J_1-J_2)Q_{10,t}(x,t)-i(I_1-I_2)Q_{10,x}(x,t)
-k\epsilon_2 Q_{11}(x,t)Q^{\ast}_{01}(x,t)=0,\\
iJ_1Q_{11,t}(x,t)-iI_1Q_{11,x}(x,t)
-k(Q_{10}Q_{01}+\epsilon_2 Q_{12}Q^{\ast}_{01})(x,t)=0,\\
i(J_1+J_2)Q_{12,t}(x,t)-i(I_1+I_2)Q_{12,x}(x,t)
-kQ_{11}(x,t)Q_{01}(x,t)=0,\\
iJ_2Q_{01,t}(x,t)-iI_2Q_{01,x}(x,t)
-k\epsilon_1(Q^{\ast}_{11}Q_{12}+\epsilon_2Q^{\ast}_{10}Q_{11})(x,t)=0.
\end{split}\label{waves_orth}\end{equation}
The $\bbbz_2$ reduction requires a dressing factor in the form
\begin{eqnarray}\label{factor1}
u(x,t,\lambda)=\openone+\left(\frac{1}{c(\lambda)}-1\right)P(x,t)
+\left(c(\lambda)-1\right)KSP^{\ast}(x,t)(KS)^{-1},
\end{eqnarray}
where $P(x,t)$ is a projector of first rank (compare with (\ref{u_orthogon}))
\begin{equation}
P(x,t)=\frac{K|m^{\ast}(x,t)\rangle\langle m(x,t)|}
{\langle m{^\ast}(x,t)|K|m(x,t)\rangle},
\qquad \lambda^{-}=(\lambda^{+})^{\ast}
\end{equation}
and $S$ is the matrix of the metric in $\bbbc^5$ which is involved in the
definition of the orthogonal algebra $so(5)$, namely
\[so(5)=\left\{ A\in sl(5): A^TS+SA=0\right\},\qquad  S_{ij}=(-1)^{i-1}\delta_{i,6-j}.\]
The generic 1-soliton solution obtained by using the dressing method
is the following
\begin{equation}
\begin{split}
Q_{10}(z)= \frac{2i\nu}{\langle m^{\ast}|K|m\rangle}
\left(\epsilon_1 (m_{0}^1)^{\ast} m_{0}^2 e^{i\left(\lambda^{+}z_2
-{\lambda^{+}}^{\ast}z_1\right)}
+\epsilon_2 (m_{0}^4)^{\ast} m_{0}^5 e^{i\left({\lambda^{+}}^{\ast}z_2-\lambda^{+}z_1\right)}\right),\\
Q_{11}(z)=\frac{2i\nu}{\langle m^{\ast}|K|m\rangle}\left(\epsilon_1
(m_{0}^1)^{\ast} m_{0}^3 e^{-i{\lambda^{+}}^{\ast}z_1}
-(m_{0}^3)^{\ast} m_{0}^5 e^{-i\lambda^{+} z_1}\right),\\
Q_{12}(z)=\frac{2i\nu}{\langle m^{\ast}|K|m\rangle}
\left(\epsilon_1 (m_{0}^1)^{\ast} m_{0}^4
e^{-i\left({\lambda^{+}}^{\ast}z_1+\lambda^{+}z_2\right)}
+\epsilon_2 (m_{0}^2)^{\ast} m_{0}^5 e^{-i\left(\lambda^{+}z_1+{\lambda^{+}}^{\ast}z_2\right)}\right),\\
Q_{01}(z)=\frac{2i\nu}{\langle m^{\ast}|K|m\rangle}\left(\epsilon_2 (m_{0}^2)^{\ast} m_{0}^3e^{-i{\lambda^{+}}^{\ast}z_2}+(m_{0}^3)^{\ast} m_{0}^4e^{-i\lambda^{+}z_2}\right),\\
\langle m^{\ast}|K|m\rangle=\epsilon_1|m_{0}^1|^2e^{-2\nu z_1}
+\epsilon_2|m_{0}^2|^2e^{-2\nu z_2}+|m_{0}^3|^2
+\epsilon_2|m_{0}^4|^2e^{2\nu z_2}+\epsilon_1|m_{0}^5|^2e^{2\nu z_1},
\end{split}\end{equation}
where $|m_0\rangle$ is a constant vector (polarization vector).
By imposing certain resrictions on the components of $|m(x,t)\rangle$
we can obtain NLEE associated with some subalgebra of $so(5)$. Let us
consider several simple examlpes:
\begin{enumerate}
\item Suppose $m_{0}^1=m_{0}^5=0$. The only nonzero wave is $Q_{01}(x,t)$
related to the simple root $\alpha_2$ (of course, we mean an independent
nonzero wave since $Q_{-\alpha_2}$ is nonzero too). Thus we conclude that
the solution is a $sl(2)$ soliton. Another $sl(2)$ soliton is derived when
$m_{0}^2=m_{0}^4=0$ is satisfied. In this case $Q_{11}(x,t)$ is the
nonvanishing component, repectively the sl(2) subalgebra is connected
with the root $e_1=\alpha_1+\alpha_2$.

\item Let $m_{0}^3=0$ is fulfilled. Then we see that $Q_{10}(x,t)$ and
$Q_{12}(x,t)$ are nonzero waves. Since the corresponding Weyl generators
commute this determines a representation of $sl(2)\oplus sl(2)\approx so(4)$
in $so(5)$.

\item Impose the resrictions $(m_{0}^1)^{\ast}=m_{0}^5$,
$(m_{0}^2)^{\ast}=m_{0}^4$ and $(m_{0}^3)^{\ast}=m_{0}^3$.
As a result we obtain
\begin{equation}
\begin{split}
Q_{10}(z)=\frac{i\nu}{\Delta_1}\sinh 2\theta_0\cosh(\nu(z_1+z_2))e^{-i\mu(z_1-z_2+\phi_1-\phi_2)},\\
Q_{11}(z)=-\frac{2\sqrt{2}i\nu}{\Delta_1}\sinh\theta_0\sinh(\nu z_1)e^{-i(\mu z_1+\phi_1)},\\
Q_{12}(z)=\frac{i\nu}{\Delta_1}\sinh(2\theta_0)\cosh(\nu(z_1-z_2))e^{-i\mu(z_1+z_2+\phi_1+\phi_2)},\\
Q_{01}(z)=\frac{2\sqrt{2}i\nu}{\Delta_1}\cosh\theta_0\cosh(\nu z_2)e^{-i\mu(z_2+\phi_2)},
\end{split}\end{equation}
where we have used the representation
\[m_{0}^1=\frac{m_{0}^3}{\sqrt{2}}\sinh\theta_0 e^{i\phi_1},\qquad
m_{0}^2=\frac{m_{0}^3}{\sqrt{2}}\cosh\theta_0 e^{i\phi_2},\qquad \theta_0\in\bbbr,\]
\[\Delta_1(x,t)=2\left(\sinh^2\theta_0\sinh^2(\nu z_1)+\cosh^2\theta_0\cosh^2(\nu z_2)\right).\]
In particular, if $\theta_0=0$ then one obtains a single wave
\begin{equation}
Q_{01}(x,t)=\frac{\sqrt{2}i\nu}{\cosh(\nu z_2)}e^{-i\mu(z_2+\phi_2)}.
\end{equation}
\end{enumerate}

\begin{remark}
In the "soliton sector" the first two examples are trivial meaning that the 4-wave system
(\ref{waves_orth}) is linearized. However, they have a nontrivial application
when one constructs the 2-soliton soliton by the dressing 1-soliton solution
and when dressing a general FAS $\chi^{\pm}_0(x,t)$.
\end{remark}

\subsection{Doublet Solitons}\label{sec:5.1}
In this subsection we are going to derive a 1-soliton solution to a 4-wave
system with an additional $\bbbz_2$ symmetry imposed on it. This is equivalent
to a $\bbbz_2\times\bbbz_2$ symmetry condition imposed to it. Let the action
of $\mathbb{Z}_2\times\mathbb{Z}_2$ in the space of fundamental
solutions of the linear problem is given by
\[\chi^{-}(x,t,\lambda)=K_1\left((\chi^{+})^{\dag}(x,t,\lambda^{\ast})\right)^{-1}K^{-1}_1\]
\[\chi^{-}(x,t,\lambda)=K_2\left((\chi^{+})^T(x,t,-\lambda)\right)^{-1}K^{-1}_2\]
where $K_{1,2}\in SO(5)$ and $[K_1,K_2]=0$.
Consequently the potential $U(x,t,\lambda)$ satisfies the symmetry conditions (\ref{sym_pot1}) and
(\ref{sym_pot2}). The $\mathbb{Z}_2\times\mathbb{Z}_2$-reduced 4-wave system reads
\begin{equation}
\begin{split}
(J_1-J_2)\mathbf{q}_{10,t}(x,t)-(I_1-I_2)\mathbf{q}_{10,x}(x,t)
+k\mathbf{q}_{11}(x,t)\mathbf{q}_{01}(x,t)=0,\\
J_1\mathbf{q}_{11,t}(x,t)-I_1\mathbf{q}_{11,x}(x,t)
+k(\mathbf{q}_{12}(x,t)-\mathbf{q}_{10}(x,t))\mathbf{q}_{01}(x,t)=0,\\
(J_1+J_2)\mathbf{q}_{12,t}(x,t)-(I_1+I_2)\mathbf{q}_{12,x}(x,t)
-k\mathbf{q}_{11}(x,t)\mathbf{q}_{01}(x,t)=0,\\
J_2\mathbf{q}_{01,t}(x,t)- I_2\mathbf{q}_{01,x}(x,t)
+k(\mathbf{q}_{10}(x,t)+\mathbf{q}_{12}(x,t))q_{11}(x,t)=0,
\end{split}\end{equation}
where $\mathbf{q}_{10}(x,t)$, $\mathbf{q}_{11}(x,t)$, $\mathbf{q}_{12}(x,t)$ and
$\mathbf{q}_{01}(x,t)$ are real valued fields and their indices are
associated with the basis of simple roots of $\mathbf{B}_2$
introduced in the previous section, i. e.
\[Q_{10}(x,t)=i \mathbf{q}_{10}(x,t),\qquad Q_{11}(x,t)=i\mathbf{q}_{11}(x,t),
\qquad Q_{12}(x,t)=i \mathbf{q}_{12}(x,t),\qquad Q_{01}(x,t)=i\mathbf{q}_{01}(x,t).\]
The constant $k$ coincides with that one in the previous examples
\[k:=J_1I_2-J_2I_1.\]

In accordance with what we said in previous chapter the dressing
factor $g(x,t,\lambda)$ must be invariant under the action of
$\mathbb{Z}_2\times\mathbb{Z}_2$, i.e.
\begin{equation}\label{ginv_1}
K_1\left(u^{\dag}(x,t,\lambda^{\ast})\right)^{-1}K^{-1}_1=u(x,t,\lambda)
\end{equation}
\begin{equation}\label{ginv_2}
K_2\left(u^T(x,t,-\lambda)\right)^{-1}K^{-1}_2=u(x,t,\lambda).
\end{equation}
Let for the sake of simplicity require that $K_1=K_2=\openone$. As
a result we find that the poles of the dressing matrix are purely
imaginery, i.e.
\[\lambda^{\pm}=\pm i\nu,\qquad \nu>0 .\]
Thus the invariance conditions (\ref{ginv_1}) and (\ref{ginv_2}) implies that
the dressing matrix gets the form
\[u(x,t,\lambda)=\openone+\frac{2i\nu}{\lambda-i\nu}P(x,t)
-\frac{2i\nu}{\lambda+i\nu}SP^{\ast}(x,t)S, \qquad P^{\ast}(x,t)=P(x,t).\]
In the simplest case the explicit form of $P(x,t)$ is
\[P(x,t)=\frac{|m(x,t)\rangle\langle m(x,t)|}{\langle m(x,t)|m(x,t)\rangle},\]
where the vector $|m(x,t)\rangle=e^{-\nu(Jx+It)}|m_0\rangle$ is real.
Therefore the solution  turns into
\begin{equation}
\begin{split}
\mathbf{q}_{10}(x,t)=\frac{2\nu}{\langle m|m\rangle}
\left(e^{-\nu[(J_1+J_2)x+(I_1+I_2)t]}m_{0}^1 m_{0}^2
+e^{\nu[(J_1+J_2)x+(I_1+I_2)t]}m_{0}^5 m_{0}^4\right),\\
\mathbf{q}_{11}(x,t)=\frac{2\nu}{\langle m|m\rangle}
\left(e^{-\nu(J_1x+I_1t)}m_{0}^1 m_{0}^3
-e^{\nu(J_1x+I_1t)}m_{0}^5 m_{0}^3\right),\\
\mathbf{q}_{12}(x,t)=\frac{2\nu}{\langle m|m\rangle}
\left(e^{-\nu[(J_1-J_2)x+(I_1-I_2)t]}m_{0}^1 m_{0}^4
+e^{\nu[(J_1-J_2)x+(I_1-I_2)x]}m_{0}^5 m_{0}^2\right),\\
\mathbf{q}_{01}(x,t)=\frac{2\nu}{\langle m|m\rangle}
\left(e^{-\nu(J_2x+I_2t)}m_{0}^2 m_{0}^3
+e^{\nu(J_2x+I_2t)}m_{0}^4 m_{0}^3\right).
\end{split}\end{equation}
These solutions can be rewritten in terms of hyperbolic functions as follows
\begin{equation}
\begin{split}
\mathbf{q}_{10}(x,t)=\frac{4\nu}{\langle m|m\rangle}N_1N_2
\cosh \{\nu[(J_1+J_2)x+(I_1+I_2)t]+\delta_1+\delta_2\},\\
\mathbf{q}_{11}(x,t)=-\frac{4\nu}{\langle m|m\rangle}N_1m_{0}^3\sinh[\nu(J_1x+I_1t)+\delta_1],\\
\mathbf{q}_{12}(x,t)=\frac{4\nu}{\langle m|m\rangle}N_1N_2\cosh\{\nu[(J_1-J_2)x+(I_1-I_2)t]
+\delta_1-\delta_2\},\\
\mathbf{q}_{01}(x,t)=\frac{4\nu}{\langle m|m\rangle}N_2m_{0}^3\cosh[\nu(J_2x+I_2t)+\delta_2],
\end{split}\end{equation}
\[\langle m(x,t)|m(x,t)\rangle=2N^2_1\cosh 2(\nu(J_1x+I_1t)+\delta_1)+2N^2_2\cosh 2(\nu(J_2x+I_2t)+\delta_2)+(m_{0}^3)^2,\]
where we have implied that $m_{0}^k>0$ for $k=1,2,4,5$ and therefore the following expressions
\[\delta_1:=\frac{1}{2}\ln\frac{m_{0}^5}{m_{0}^1},\qquad\delta_2:=\frac{1}{2}\ln\frac{m_{0}^4}{m_{0}^2},
\qquad N_1=\sqrt{m_{0}^1m_{0}^5},\qquad N_2=\sqrt{m_{0}^2 m_{0}^4}.\]
make some sense.

In particular, when $m_{0}^1=m_{0}^5$ and $m_{0}^2=m_{0}^4$ or in other words $\delta_1=\delta_2=0$
we obtain
\[\mathbf{q}_{10}(x,t)=\frac{\nu}{\Delta_{D}}\sinh(2\theta_0)\cosh\nu[(J_1+J_2)x+(I_1+I_2)t],\]
\[\mathbf{q}_{11}(x,t)=-\frac{2\sqrt{2}\nu}{\Delta_D}\sinh\theta_0\sinh\nu(J_1x+I_1t),\]
\[\mathbf{q}_{12}(x,t)=\frac{\nu}{\Delta_D}\sinh(2\theta_0)\cosh\nu[(J_1-J_2)x+(I_1-I_2)t],\]
\[\mathbf{q}_{01}(x,t)=\frac{2\sqrt{2}\nu}{\Delta_D}\cosh\theta_0\cosh\nu(J_2x+I_2t).\]
where $\theta_0\in\mathbb{R}$ and
\[ \Delta_{D}(x,t):=2\left(\sinh^2\theta_0\sinh^2\nu(J_1x+I_1t)+\cosh^2\theta_0\cosh^2\nu(J_2x+I_2t)\right).\]

\subsection{Quadruplet Solitons}\label{sec:5.2}

This time to ensure the proper $\bbbz_2\times\bbbz_2$-invariance of $u(x,t,\lambda)$
we add in the expression for it two terms more. The requirements
(\ref{ginv_1})--(\ref{ginv_2}) lead to the following dressing matrix
\begin{equation}\label{factor}
\begin{split}
u(x,t,\lambda)& = \openone+\frac{A(x,t)}{\lambda-\lambda^{+}}
+\frac{K_1SA^{\ast}(x,t)(K_1S)^{-1}}{\lambda-(\lambda^{+})^{\ast}}
 -\frac{K_2SA(x,t)(K_2S)^{-1}}{\lambda+\lambda^{+}}\\
& -\frac{K_1K_2A^{\ast}(x,t)(K_1K_2)^{-1}}{\lambda+(\lambda^{+})^{\ast}}.
\end{split}
\end{equation}

Taking into account the explicit formula (\ref{factor}) one can derive
in the soliton case $Q_0(x,t)\equiv 0$ the following relation
\begin{equation}\label{breather}
[J,Q](x,t)=[J,A+K_1SA^{\ast}SK_1-K_2SASK_2-K_1K_2A^{\ast}K_2K_1](x,t).
\end{equation}
Like in previous considerations we decompose the matrix $A(x,t)$ using two
matrix factors $|X(x,t)\rangle$ and $|F(x,t)\rangle$ and derive some differential equation
for $|F(x,t)\rangle$ which leads to
\[|F(x,t)\rangle=e^{i\lambda^{+}(Jx+It)}|F_0\rangle.\]
The linear system for $|X(x,t)\rangle$ in this case is following
\[\left(\openone-\frac{K_2SA(x,t)SK_2}{2\lambda^{+}}+\frac{K_1SA^{\ast}(x,t)SK_1}
{2 i\nu}-\frac{K_1K_2A^{\ast}(x,t)K_2K_1}{2\mu}\right)S|F(x,t)\rangle=0.\]
Starting from the above mentioned equation we can derive the following
auxilliary linear system
\[S|F(x,t)\rangle=\frac{\langle G|S|F\rangle}{2\lambda^{+}}|Y\rangle
-\frac{\langle H|S|F\rangle}{2i\nu}|Z\rangle+\frac{\langle N|S|F\rangle}{2\mu}|W\rangle\]
\[S|G(x,t)\rangle=\frac{\langle F|S|G\rangle}{2\lambda^{+}}|X\rangle
+\frac{\langle H|S|G\rangle}{2\mu}|Z\rangle-\frac{\langle N|S|G\rangle}{2i\nu}|W\rangle\]
\[S|H(x,t)\rangle=\frac{\langle F|S|H\rangle}{2i\nu}|X\rangle
+\frac{\langle G|S|H\rangle}{2\mu}|Y\rangle+ \frac{\langle N|S|H\rangle)}{2(\lambda^{+})^{\ast}}|W\rangle\]
\[S|N(x,t)\rangle=\frac{\langle F|S|N\rangle}{2\mu}|X\rangle
+\frac{\langle G|S|N\rangle)}{2i\nu}|Y\rangle+\frac{\langle H|S|N\rangle}{2(\lambda^{+})^{\ast}}|Z\rangle\]
where we introduced auxiliary entities
\[|Y(x,t)\rangle:=K_2S|X(x,t)\rangle,\qquad |Z(x,t)\rangle:=K_1S|X^{\ast}\rangle(x,t),\qquad |W(x,t)\rangle:=K_1K_2|X^{\ast}(x,t)\rangle,\]
\[|G(x,t)\rangle:=K_2S|F(x,t)\rangle,\quad |H(x,t)\rangle:=K_1S|F^{\ast}(x,t)\rangle,\quad
|N(x,t)\rangle:=K_1K_2|F^{\ast}(x,t)\rangle.\]
To calculate $|X(x,t)\rangle$ we just have to solve
the linear system shown above. The answer reads
\[\left(\begin{array}{c} |X\rangle\\ |Y\rangle\\ |Z\rangle\\ |W\rangle\end{array}\right)(x,t)=
\frac{1}{\Delta(x,t)}\left(\begin{array}{cccc}
0 & a^{\ast} & b & -c\\
a^{\ast} & 0 & -c & b\\
-b & -c & 0 & a\\
-c & -b & a & 0\end{array}\right)\left(\begin{array}{c} S|F\rangle\\
S|G\rangle\\S|H\rangle\\S|N\rangle\end{array}\right)(x,t),\]
where
\[\Delta(x,t):=|a(x,t)|^2+b^2(x,t)-c^2(x,t),\qquad
a(x,t):=\frac{\langle F(x,t)|S|G(x,t)\rangle}{2\lambda^{+}},\]
\[b(x,t):=\frac{\langle F(x,t)|S|H(x,t)\rangle}{2i\nu},\qquad
c(x,t):=\frac{\langle F(x,t)|S|N(x,t)\rangle}{2\mu}.\]
Finally putting the result for $|X(x,t)\rangle$ in (\ref{breather}) we obtain
the generic quadruplet solution to the 4-wave system associated with the $\mathbf{B}_2$
algebra (suppose $K_1=K_2=\openone$)
\begin{eqnarray*}
\mathbf{q}_{10}(x,t) & = & \frac{4}{\Delta}\im\left[a^{\ast}N_1
\cosh(\varphi_1+\varphi_2)-\frac{imN_1^{\ast}}{\mu\nu}\left(\mu\cosh(\varphi^{\ast}_1+\varphi_2)
-i\nu\cosh(\varphi^{\ast}_1-\varphi_2)\right)\right]N_2\\
\mathbf{q}_{11}(x,t) & = &
\frac{4}{\Delta}\im\left[a^{\ast}N_1\sinh(\varphi_1)
-\frac{im\lambda^{+}}{\mu\nu}N^{\ast}_1\sinh(\varphi^{\ast}_1)\right]m_{0}^3\\
\mathbf{q}_{12}(x,t) & = &
\frac{4}{\Delta}\im\left[a^{\ast}N_1\cosh(\varphi_1-\varphi_2)
-\frac{im N^{\ast}_1}{\mu\nu}\left(\mu\cosh(\varphi^{\ast}_1-\varphi_2)
-i\nu\cosh(\varphi^{\ast}_1+\varphi_2)\right)\right]N_2\\
\mathbf{q}_{01}(x,t) & = &
\frac{4}{\Delta}\im\left[a^{\ast}N_2\cosh(\varphi_2)
-\frac{im{\lambda^{+}}^{\ast}}{\mu\nu}N^{\ast}_2\cosh(\varphi^{\ast}_2)\right]m_{0}^3.
\end{eqnarray*}
where
\[a(x,t)=\frac{1}{\mu+i\nu}\left[N^2_1\cosh 2\varphi_1+N^2_2\cosh 2\varphi_2
+\frac{F^2_{0,3}}{2}\right],\qquad b(x,t)=\frac{m(x,t)}{i\nu},\qquad
c(x,t)=\frac{m(x,t)}{\mu},\]
\[m(x,t)=|N_1|^2\cosh (2\mbox{Re}\ \varphi_1)+|N_2|^2\cosh (2\mbox{Re}\ \varphi_2)+\frac{|m_{0}^3|^2}{2},\qquad
N_{\sigma}:=\sqrt{m_{0}^{\sigma}m_{0}^{6-\sigma}},\]
\[\varphi_{\sigma}(x,t):=i\lambda^{+}(J_{\sigma}x+I_{\sigma}t)
+\frac{1}{2}\log\frac{m_{0}^{\sigma}}{m_{0}^{6-\sigma}},
\qquad\sigma=1,2.\]

\section{Discussion and further studies}\label{sec:FS}

Here we shall outline some further topics which could be studied and which could
lead to a deeper understanding of these soliton properties.

The first obvious remark is that $sl(n) $ contains as subalgebras also
$so(p) $ and $sp(p) $ subalgebras. So it will be interesting to specify
the conditions under which $L(\lambda ) $ has solitons of type $so(p) $ or
$sp(p) $.

Second remark of the same nature is that one can start with
$L(\lambda ) $ related to $so(n) $ or $sp(n) $ algebras; such
generalized Zakharov-Shabat systems allow one to solve special
types of $N $-wave systems whose soliton solutions have not yet
been classified. Such systems, due to the additional symmetry,
have a richer structure.

The explicit form of the corresponding $N $-wave system  related to these
algebras has been reported in \cite{ForKu,ZaMi,Harnad1}, see also
\cite{59,RIa}.  What could be done is to analyze the structure of its
soliton solutions \cite{59,RIa} which are more involved due to the
additional orthogonal symmetry involved.  However this symmetry complicates
the construction of the dressing factors.  Nevertheless, interesting new types of
integrable cubic interactions could be obtained.

Also, even more complicated types of solitons will be related to
projectors of higher rank. The projectors $P(x,t) $, which we used above,
were all of rank 1. The rank 2 projector $P_2 $ can be defined as:
\begin{equation}\label{eq:P_2}
P_2(x,t) = \sum_{ks}^{2} |n_k\rangle \hat{M}_{ks}(x,t)
\langle n_s^\ast  |, \qquad  M_{sk}(x,t) = \langle n_s^\ast
|n_k\rangle, \qquad \hat{M} \equiv M^{-1}.
\end{equation}
Now each soliton will be parametrized by two polarization vectors;
the corresponding eigensubspace will be two-dimensional too. Among
the various types of rank-2 one-soliton solutions, there will be
various possible configurations for the two polarization vectors.
An example of a dressing factor $u(x,t,\lambda)$ constructed by
a projector of second rank is the following one
\[u(x,t,\lambda)=\openone+\left(\frac{1}{\sqrt{c(\lambda)}}-1\right)P_2
+(\sqrt{c(\lambda)}-1)\overline{P}_2.\]
Such type of a dressing factor was used by Wadati and coauthors \cite{wad}
to derive the soliton solutions to a multicomponent Schr\"odinger equation
relate to symplectic algebra $sp(4)$.

It is known in general how  the machinery, well understood for the AKNS
system such as Wronskian relations, expansions over `squared solutions',
etc. can be generalized also for these types of systems. The dressing
method, after some modifications, can also be applied, leading to the
derivation of their soliton solutions.

An interesting problem is the study of how the different possible
reductions (see e.g.  \cite{59}) of these systems will influence the
number of one-soliton types.

Soliton interactions for various different types of solitons of these
systems also present interesting problems.
>From the results known for the $N $-wave systems \cite{ZaMa,DJK0} it is known
that new effects in soliton interaction, such as soliton decay and
soliton fusion may arise.

\section*{Acknowledgements}

This research has been supported in part by the USA National Science
Foundation and the USA Air Force Office of Scientific Research.

\end{document}